\documentclass[letterpaper, 10pt, conference]{ieeeconf} 

\IEEEoverridecommandlockouts                              

\usepackage[noadjust]{cite}
\usepackage{graphics} 
\usepackage{epsfig} 
\usepackage{times} 
\usepackage{amssymb, amsmath}
\usepackage[font=small,labelfont=bf]{caption}
\usepackage{subcaption}
\usepackage{breqn}
\usepackage{array}
\usepackage{float}
\usepackage{siunitx}
\usepackage{mathtools}
\usepackage{float}
\usepackage{siunitx}

\usepackage{amsthm}
\usepackage{xspace}
\usepackage[linesnumbered,ruled,vlined]{algorithm2e}
\usepackage{array}
\usepackage{algorithmic}
\usepackage{makecell}
\usepackage{booktabs, caption, makecell}

\usepackage{threeparttable}
\usepackage[symbol]{footmisc}

\makeatletter
\DeclareRobustCommand\onedot{\futurelet\@let@token\@onedot}
\def\@onedot{\ifx\@let@token.\else.\null\fi\xspace}

\def\eg{\emph{e.g}\onedot} 
\def\ie{i.e.}

\makeatother

\newtheorem{property}{Property}

\theoremstyle{definition}
\newtheorem{definition}{Definition}
\theoremstyle{remark}

\theoremstyle{definition}

\theoremstyle{definition}

\DeclareMathOperator{\MSE}{MSE}

\DeclareMathOperator*{\esssup}{ess\,sup}

\renewcommand{\cal}[1]{\mathcal{ #1 }}

\newcommand{\bb}[1]{\mathbb{ #1 }}

\newcommand{\grad}{\nabla}
\newcommand{\intersect}{\cap}

\newcommand{\R}{\bb{R}}

\DeclarePairedDelimiter\floor{\lfloor}{\rfloor}
\DeclarePairedDelimiter\abs{\lvert}{\rvert}%
\DeclarePairedDelimiter\norm{\lVert}{\rVert}%

\DeclarePairedDelimiterX{\Set}[2]\{\}{%
  \, #1 \;\delimsize\vert\; #2 \,
}

\title{\LARGE \bf
Neural Lyapunov Control for Nonlinear Systems\\ with Unstructured Uncertainties
}

\author{Shiqing Wei, Prashanth Krishnamurthy, and Farshad Khorrami
\thanks{The authors are with Control/Robotics Research Laboratory, Dept. of ECE, NYU Tandon School of Engineering, 5 Metrotech Center, Brooklyn, NY 11201, USA. 
{\tt\small \{shiqing.wei, prashanth.krishnamurthy, khorrami\}@nyu.edu}}%
\thanks{This work was supported in part by ARO grant  W911NF-22-1-0028 and in part by the New York University Abu Dhabi (NYUAD) Center for Artificial Intelligence and Robotics, funded by Tamkeen under the NYUAD Research Institute Award CG010.}
}

\begin{document}

\maketitle
\thispagestyle{empty}
\pagestyle{empty}

\begin{abstract}
Stabilizing controller design and region of attraction (RoA) estimation are essential in nonlinear control. Moreover, it is challenging to implement a control Lyapunov function (CLF) in practice when only partial knowledge of the system is available. We propose a learning framework that can synthesize state-feedback controllers and a CLF for control-affine nonlinear systems with unstructured uncertainties. Based on a regularity condition on these uncertainties, we model them as bounded disturbances and prove that a CLF for the nominal system (estimate of the true system) is an input-to-state stable control Lyapunov function (ISS-CLF) for the true system when the CLF's gradient is bounded. We integrate the robust Lyapunov analysis with the learning of both the control law and CLF. We demonstrate the effectiveness of our learning framework on several examples, such as an inverted pendulum system, a strict-feedback system, and a cart-pole system.
\end{abstract}

\section{Introduction}
While the knowledge of a control Lyapunov function (CLF) for a system can enable the implementation of universal controllers \cite{sontag1989universal}, designing a CLF for a given real-world nonlinear system can be highly challenging, especially when the system's dynamics are uncertain. Furthermore, when faced with a real-world uncertain system, proving that a controller provides global stabilization might be infeasible or even ill-defined, and one instead needs to estimate the region of attraction (RoA) of the closed-loop system using online data. Motivated by the several advances in learning-based methods for various control design tasks (\eg, \cite{richards2018lyapunov,dai2021state, DBLP:conf/cdc/DaiKK22, mehrjou2021neural, wei2020towards}), we consider the problems of designing a controller for uncertain systems and estimating the achieved RoA from a learning-based or data-driven approach in this paper. 

Specifically, we address the problem of synthesizing state-feedback controllers and estimating the RoA for control-affine nonlinear systems with unstructured uncertainties. We model these uncertainties as a bounded disturbance but with no assumptions about the source of these uncertainties. They may result from incorrect model parameters or dynamics that are not reflected in the model. Our work takes advantage of the rich literature on Lyapunov theory \cite{lyapunov1992general, artstein1983stabilization, sontag1995characterizations,KK06}, and a data-driven approach is adopted. 

Several approaches have been explored in the existing literature to enhance the robustness of Lyapunov analysis and control designs to uncertainties in the underlying dynamics of real-world systems, such as an unknown constant parameter \cite{krstic1995control}, an unknown Gaussian process \cite{berkenkamp2016safe}, or an unknown linearly parameterized control-affine system \cite{sun2021lyapunov}. In particular, the authors of \cite{taylor2019control} consider the modeling error as a disturbance and propose the projection-to-state stability approach to characterize the tracking error. However, one common point among the above methods is that they either derive controllers \cite{krstic1995control, taylor2019control, sun2021lyapunov} or refine the RoA \cite{berkenkamp2016safe} based on a given CLF. In practice, such a CLF may not be readily available for complex nonlinear systems and may provide a conservative estimation of the actual RoA.

Another area closely related to this paper is the automated formulation of Lyapunov functions. Lyapunov functions for polynomial systems can be found by solving linear matrix inequalities (LMIs) \cite{tibken2000estimation}. Approximation of Lyapunov functions by sum-of-squares (SOS) polynomials can be found through the solution of a semidefinite programming (SDP) problem \cite{papachristodoulou2002construction}. Computational methods for Lyapunov functions have been reviewed in \cite{giesl2015review}. More recently, neural networks have been used to approximate a Lyapunov function with SMT (Satisfiability Modulo Theories) solvers being a verification tool \cite{chang2019neural, zhao2021neural}. However, these learning-based approaches (\eg, \cite{chang2019neural, mehrjou2021neural}) have been developed assuming exact knowledge of the system. In \cite{richards2018lyapunov}, knowledge of closed-loop dynamics is needed when computing the Lipschitz constant. The method proposed in \cite{mehrjou2021neural}, Neural Lyapunov Redesign (NLR), is an offline method that finds a neural Lyapunov function and a stabilizing controller assuming known system dynamics. 

\textit{Our contributions:} We consider control-affine nonlinear systems with unstructured uncertainties and propose a learning framework that simultaneously learns the following: a state-feedback controller that seeks to enlarge the RoA, a CLF that can be used to estimate the RoA, and an improved model of the uncertain system dynamics (starting with an initial nominal approximate system model).  Inspired by \cite{taylor2019control}, we model the unstructured uncertainties as a bounded disturbance. We prove that when the gradient of the CLF is bounded, and the disturbance is bounded by a particular quantity, a CLF for the nominal system is an input-to-state stable control Lyapunov function (ISS-CLF) for the true system and thus can correctly estimate the RoA for the true system. We apply a machine learning-based data-driven approach for learning the controller, CLF, and system dynamics and demonstrate the effectiveness of our learning framework on three different examples.

\section{Preliminaries} \label{sec:preliminary}
Let $\cal{X} \subset \R^n$ and $\cal{U} \subset \R^m$ be the state and control input spaces (in general, subsets of $\R^n$ and $\R^m$ to model physical constraints of real-world systems). Consider the following control-affine system:
\begin{equation}\label{eq:true_sys}
    \dot{x} = f(x) + g(x)u
\end{equation}
with drift dynamics $f: \cal{X} \to \R^n$ and actuation matrix $g: \cal{X} \to \R^{n \times m}$. To ensure the existence and uniqueness of the solution, we assume that $f$ and $g$ are Lipschitz continuous on $\cal{X}$. We further assume $0 \in \cal{X}$ and $f(0) + g(0)u_0 = 0$ for a certain $u_0 \in \cal{U}$ and also assume controllability of the system. Introducing a disturbance in \eqref{eq:true_sys}, we consider the perturbed system:
\begin{equation}\label{eq:disturbed_sys}
    \dot{x} = f(x) + g(x)u + d
\end{equation}
where $d \in \cal{D}$ is the disturbance assumed to be essentially bounded in time (\ie, bounded everywhere except possibly on a set of measure zero) and $\cal{D} \subset \R^n$ is the disturbance space. A natural framework for modeling the effects of perturbations is given by the widely used notions of input-to-state stability (ISS) \cite{sontag1989smooth} and input-to-state stable control Lyapunov functions (ISS-CLF) \cite{sontag1995characterizations}. 

\begin{definition}[\it{CLF and ISS-CLF}]\label{def:clf_issclf}
A class $C^1$ function $V: \cal{X} \to \R_+$ is a CLF for $(\ref{eq:true_sys})$ on $\cal{X}$ if there exist $\alpha_1, \alpha_2, \alpha_3 \in \cal{K}_{\infty}$ such that for all $x \in \cal{X}$:
\begin{align}
    \alpha_1 (\norm{x}) \leq V(x) &\leq \alpha_2 (\norm{x})\label{eq:bound},\\
    \inf_{u \in \cal{U}} \dot{V}(x, u) &\leq -\alpha_3(\norm{x}) \label{eq:condition_clf}.
\end{align}
$V$ is an ISS-CLF for (\ref{eq:disturbed_sys}) on $\cal{X}$ if it satisfies \eqref{eq:bound} and additionally, there exist $\alpha_4, \rho \in \cal{K}_{\infty}$ such that \begin{equation}\label{eq:condition_issclf}
    \norm{x} \geq \rho\left(\esssup_{\tau \geq t_0} \norm{d(\tau)} \right) \Rightarrow \inf_{u \in \cal{U}} \dot{V}(x, u, d) \leq -\alpha_4(\norm{x})
\end{equation}
for all $x \in \cal{X}$ and $d \in \cal{D}$. 
\end{definition}

The existence of a CLF implies the existence of a state-feedback controller $k: \cal{X} \rightarrow \cal{U}$ \cite{artstein1983stabilization} such that $\dot V \leq -\alpha_3(||x||)$ in the closed-loop system (i.e., when control inputs are generated using $k$). We refer to controllers satisfying this inequality as \textit{admissible}.
We note that for the definition of a CLF, $\alpha_1$ and $\alpha_2$ need only to be class $\cal{K}$ functions, and condition \eqref{eq:condition_clf} can be reduced to $\inf_{u \in \cal{U}} \dot{V}(x, u) < 0$ to guarantee local asymptotic stability of the origin \cite{artstein1983stabilization}. 

\section{Uncertain Dynamics} \label{sec:method}
Let $\hat{f}: \cal{X} \to \R^n$ and $\hat{g}: \cal{X} \to \R^{n \times m}$ be Lipschitz continuous functions denoting the estimates of $f$ and $g$ in \eqref{eq:true_sys}. Then, system \eqref{eq:true_sys}, called \textit{true system}, can be written as 
\begin{equation}\label{eq:decomp_sys}
    \dot{x} = \hat{f}(x) + \hat{g}(x)u + \underbrace{(f(x)-\hat{f}(x)) + (g(x)-\hat{g}(x)) u}_{d}
\end{equation}
where the estimation errors are written as the disturbance signal $d$. If $d$ is essentially bounded in time, system \eqref{eq:decomp_sys} can be seen as the perturbed system of the \textit{nominal system}
\begin{equation} \label{eq:nominal_sys}
    \dot{x} = \hat{f}(x) + \hat{g}(x)u.
\end{equation}

Finding a CLF for \eqref{eq:true_sys} valid on the entire state space $\cal{X}$ is difficult for nonlinear systems, and the conditions \eqref{eq:bound} and \eqref{eq:condition_clf} are usually satisfied only on a compact subset $\cal{C}$ of $\cal{X}$. Then, given a continuous state-feedback controller $k: \cal{X} \rightarrow \cal{U}$, the estimation error $d$ is indeed bounded on $\cal{C}$ as a result of the continuity of $f, \hat{f}, g, \hat{g}$, and $k$ on $\cal{X}$.

Given a CLF for the nominal system \eqref{eq:nominal_sys} on $\cal{C}$, by \eqref{eq:decomp_sys}, the time derivative of $V$ can be written as
\begin{equation}\label{eq:dyn_V}
    \dot{V}(x, k(x), d) = \underbrace{\grad{V}(x)^\top ( \hat{f}(x) + \hat{g}(x)k(x) )}_{\hat{\dot{V}}(x, k(x))} + \underbrace{\grad V(x)^\top d}_{\delta}
\end{equation}
where $\grad V$ is the gradient of $V$ and $\delta = \grad V(x)^\top d$ can be seen as a disturbance in the dynamics of $V$. Since $V$ is $C^1$ on $\cal{X}$, $\norm{\grad V}$ is bounded on $\cal{C}$ (with bound $L_V >0$), and the disturbance term $\delta$ is also bounded. 

\begin{property} \label{thm:equiv}
Let $V$ be a CLF with $\alpha_1, \alpha_2, \alpha_3 \in \cal{K}_{\infty}$ for system \eqref{eq:nominal_sys} associated with an admissible continuous controller $k: \cal{X} \rightarrow \cal{U}$ over a compact set $\cal{C} \subset \cal{X}$. Let $\Omega \subset \cal{C}$ be a sublevel set of $V$, and $\alpha_p$ and $\alpha_q$ be class $\cal{K}_\infty$ functions such that $\alpha_p + \alpha_q = \alpha_3$. If $\norm{x} \geq \alpha_q^{-1}\left(L_V\max_{x \in \Omega}\norm{d}\right)$ for all $x \in \partial \Omega$ (the boundary of set $\Omega$), then $V$ is an ISS-CLF for system \eqref{eq:decomp_sys} with controller $k$ over $\Omega$.
\end{property}


\begin{proof}
Let $c = V(x)$ for any $x \in \partial \Omega$. Then, for all $x \in \partial \Omega$, \begin{equation*}
    \dot{V}(x, k(x), d) \leq -\alpha_p(\norm{x}) - \alpha_q(\norm{x}) + \delta \leq -\alpha_p(\norm{x}) < 0,
\end{equation*} 
and by Nagumo's Theorem, $V(x(t)) \in [0,c]$ for $t \geq t_0$ if $V(x(t_0)) \in [0,c]$. Therefore, the set $\Omega$ is forward invariant, which means that for any trajectory starting within $\Omega$, the disturbance along the trajectory will be bounded by $\max_{x \in \Omega}\norm{d}$. Note that the inverse of a class $\cal{K}_\infty$ function is defined on $\R_+$ and belongs to class $\cal{K}_\infty$, $\rho: r \mapsto \alpha_q^{-1}(L_V r)$ is therefore a class $\cal{K}_\infty$ function. $V$ is an ISS-CLF on $\Omega$ since if $\norm{x} \geq \rho \left(\sup_{\tau \geq t_0} \norm{d(\tau)}\right)$, by \eqref{eq:condition_clf} and \eqref{eq:dyn_V},
\begin{equation*}
    \dot{V}(x, k(x), d) \leq -\alpha_p(\norm{x}) - \alpha_q(\norm{x}) + \delta \leq -\alpha_p(\norm{x})
\end{equation*} 
where $\alpha_p \in \cal{K}_\infty$, and condition \eqref{eq:condition_issclf} is satisfied.
\end{proof}

The next property is a direct result of Nagumo's Theorem.
\begin{property} \label{thm:forward_invariant}
Let $V$ be an ISS-CLF for \eqref{eq:decomp_sys} under a state-feedback controller $k: \cal{X} \rightarrow \cal{U}$ on a compact set $\cal{C}$ and $\Omega \subseteq \cal{C}$ be a sublevel set of $V$. $\Omega$ is forward invariant if $\norm{x} \geq \rho\left(\norm{d} \right)$ for all $x \in \partial \Omega$. 
\end{property}

Finally, based on \cite[Theorem 1]{sontag1995characterizations}, we have:
\begin{property} \label{thm:iss}
If $V$ is an ISS-CLF for system \eqref{eq:decomp_sys} under an admissible state-feedback controller $k: \cal{X} \rightarrow \cal{U}$ on $\cal{C}$, then \eqref{eq:decomp_sys} is ISS with the controller $k$ on $\cal{C}$.
\end{property}

\section{Learning the Controller and ISS-CLF}\label{sec:learning}

\subsection{Neural Network Structures}
As part of our learning framework, we update the nominal dynamics during training. For this purpose, we decompose the estimates $\hat{f}$ and $\hat{g}$ into two parts:
\begin{equation}
    \hat{f} = f_0 + f_{\theta_1}, \quad \hat{g} = g_0 + g_{\theta_2}
\end{equation}
where $f_0$ and $g_0$ depend on the initial knowledge of the system, and the residual parts $f_{\theta_1}$ and $g_{\theta_2}$ are neural networks (parameterized by ${\theta_1}$ and ${\theta_2}$, respectively). If the system is completely unknown, we may start with $f_0 =g_0=0$. 

Let $V_{\theta_3}: \cal{X} \rightarrow \R_+$, called \textit{Lyapunov candidate}, be a neural network parameterized by $\theta_3$. To ensure that $V_{\theta_3}$ satisfies  \eqref{eq:bound} in Definition~\ref{def:clf_issclf}, its structure is defined as:
\begin{equation} \label{eq:lyp_structure}
    V_{\theta_3}(x) = x^\top (M M^\top + \gamma I) x + \phi(x)^\top \phi(x)
\end{equation}
where $M \in \R^{n \times n}$ is a trainable lower triangular matrix, $I \in \R^{n \times n}$ is the identity matrix, $\gamma > 0$ is a constant and $\phi: \cal{X} \rightarrow \R^d$ is a neural network. $\theta_3$ includes both $M$ and weights in $\phi$. Following \cite{richards2018lyapunov}, $\phi$ is a composition of linear layers (with no bias term) and activation functions (Lipschitz continuous functions with a trivial null space), and its layer dimensions are increasing. Let $d_\ell$ and $W_\ell$ be the output dimension and layer weights of layer $\ell$, and
\begin{equation}\label{eq:layer_weights}
    W_\ell = \begin{bmatrix} G_{\ell 1}^\top G_{\ell 1} + \varepsilon I_{d_{\ell-1}} \\ G_{\ell 2}\end{bmatrix}
\end{equation}
where $G_{\ell 1} \in \R^{q_\ell \times d_{\ell-1}}$ for some integer $q_\ell \geq 1$, $G_{\ell 2} \in \R^{(d_\ell - d_{\ell-1}) \times d_{\ell-1}}$, $I_{d_{\ell-1}} \in \R^{d_{\ell-1} \times d_{\ell-1}}$ is the identity matrix, and $\varepsilon >0$ is a constant. Note that \eqref{eq:layer_weights} is enforced during training to ensure that $\phi$ has a trivial null space, implying that $\phi(x)^\top \phi(x)$ is positive definite. Our Lyapunov candidate differs from that of \cite{richards2018lyapunov} in two aspects. Firstly, the $\phi(x)^\top \phi(x)$ term is not necessarily lower bounded by a class $\cal{K}_\infty$ function; so we have added the $\gamma x^\top x$ term to achieve this property. Secondly, we added the $x^\top M M^\top x$ term to better capture quadratic behaviors and expect the $\phi(x)^\top \phi(x)$ term to capture other nonlinear behaviors. The upper bounding function $\alpha_2 \in \cal{K}_\infty$ can be constructed based on $V_{\theta_3}$.

The state-feedback controller $u_{\theta_4}: \cal{X} \rightarrow \cal{U}$ is also a neural network (parameterized by $\theta_4$). The activation functions of $u_{\theta_4}$ are chosen such that it is continuous.

\subsection{Learning Algorithm}
Let $\mathcal{X}_\tau \subset \mathcal{X}$ be a discretization of $\mathcal{X}$ with $\norm{x - [x]_\tau}_2 \leq \tau/2$, where $[x]_\tau$ denotes the closest point in $\mathcal{X}_\tau$ to $x \in \mathcal{X}$. Based on the structure and our choice of activation functions of the neural networks, the estimated time derivative of $V_{\theta_3}$
\begin{equation} \label{eq:est_lyp_der}
    \hat{\dot{V}}_{\theta_3}(x, u_{\theta_4}(x)) = \grad V_{\theta_3}(x) ^\top (\hat{f}(x)+\hat{g}(x)u_{\theta_4}(x))
\end{equation}
is locally Lipschitz continuous on $\cal{X}$. For $V_{\theta_3}$ to be a CLF for the nominal system \eqref{eq:nominal_sys}, condition \eqref{eq:condition_clf} in Definition~\ref{def:clf_issclf} has to be satisfied on a certain set $\cal{C}$ (to be determined), \ie, 
\begin{equation}\label{eq:condition_to_fit}
    \hat{\dot{V}}_{\theta_3}(x, u_{\theta_4}(x)) \leq -\kappa \norm{x}^2 \text{ on } \cal{C}
\end{equation}
where we have chosen $\alpha_3(\norm{x}) = \kappa \norm{x}^2$ and $\kappa$ is a positive constant. The \textit{Lyapunov loss} is defined as
\begin{align}\label{eq:training_objective}
    \mathcal{L}_{\theta_3, \theta_4} &= 
    \frac{\lambda_{\text{RoA}}}{N_i} \sum_{x \in \cal{S}_i} \text{ReLU}[\hat{\dot{V}}_{\theta_3}(x, u_{\theta_4}(x)) + \kappa \norm{x}^2 + \epsilon] \nonumber \\
    &+ \frac{\lambda_{\text{Lip}}}{N_i} \sum_{x \in \cal{S}_i} \norm{\nabla V_{\theta_3}(x)}
\end{align}
where $\lambda_{\text{RoA}}, \lambda_{\text{Lip}},$ and $\epsilon$ are positive constants, $\cal{S}_i$ is the training set at iteration $i$, $N_i$ is the number of training samples in $\cal{S}_i$, and $\text{ReLU}(x) = \max(0,x)$ stands for the Rectified Linear Unit. The first term of \eqref{eq:training_objective} accounts for condition~\eqref{eq:condition_to_fit}, and $\epsilon$ is a positive constant offset. The second term aims to limit the norm of $\grad V_{\theta_3}$. Compared with \cite{richards2018lyapunov}, our loss function has a more general structure with a regularization term that is designed to limit $L_{V_{\theta_3}}$ (upper bound of $\norm{\grad V_{\theta_3}}$), as required by part of our methodology.

\begin{algorithm}[t]
\SetAlgoLined
\KwIn{Dynamics network $f_{\theta_1}$ and $g_{\theta_2}$, Lyapunov candidate $V_{\theta_3}$, state-feedback controller $u_{\theta_4}$, loss function $\cal{L}$, mesh $\cal{X}_\tau$, initial stable set $\cal{X}_{0}$\footnotemark[2], learning rate scheduler $\text{StepLR}$, level multiplier $\eta_0$ and step $k_\eta$, and number of iterations $N$.} 

Pretrain $\hat{\dot{V}}_{\theta_3}$ on $\cal{X}_{0}$ by minimizing $\MSE(\hat{\dot{V}}_{\theta_3},\tilde{\dot{V}}_{\theta_3})$ via stochastic gradient descent (SGD)\;

\For{$i = 1, 2, ..., N$}{
    Determine $\cal{X}_{\text{stable}}$ from sampled trajectories\;
    
    $c_i = \max_{c>0} c,$ s.t. $\hat{\dot{V}}_{\theta_3}(x, u_{\theta_4}(x))\leq -\kappa \norm{x}^2$ for all $x \in \cal{X}_{\text{stable}} \intersect \cal{V}(c)$ and $\cal{V}(c) \subset \cal{X}$\;

    $\eta_i = 1 + \frac{\eta_0}{1+ \floor*{i/k_\eta}}$\;
    
    $\cal{S}_i = \cal{V}(\eta_i c_i)$\;
    
    Update $f_{\theta_1}$ and $g_{\theta_2}$ by minimizing $\MSE(\hat{\dot{V}}_{\theta_3},\tilde{\dot{V}}_{\theta_3})$ on $\cal{S}_i$ via SGD\;
    
    Determine learning rate for $\theta_3$ and $\theta_4$ by $\text{StepLR}$\;
    
    Update $V_{\theta_3}$ and $u_{\theta_4}$ by minimizing the loss function \eqref{eq:training_objective} on $\cal{S}_i$\ via SGD;
}
\caption{Learning to maximize the RoA}
\label{algo:max_roa}
\end{algorithm}

\footnotetext[2]{We assume that the algorithm starts with an initial locally stable controller.}

Our learning framework learns the nominal dynamics, the Lyapunov candidate, and the controller at the same time. Instead of directly working on full-state dynamics, we update the nominal dynamics using the dynamics of the CLF during learning. During each iteration, our approach first updates $f_{\theta_1}$ and $g_{\theta_2}$ by minimizing the mean squared error (MSE) between $\hat{\dot{V}}_{\theta_3}$ (by \eqref{eq:est_lyp_der}) and $\tilde{\dot{V}}_{\theta_3}$ (approximated value of $\dot{V}_{\theta_3}$ obtained by numerical differentiation), and then updates $V_{\theta_3}$ and $u_{\theta_4}$ by minimizing the Lyapunov loss \eqref{eq:training_objective}.

Let $\cal{V}(c) = \Set{x \in \R^n}{V_{\theta_3}(x) \leq c}$ denote the $c$-sublevel set of $V_{\theta_3}$. At iteration $i$, we collect from the system the trajectories initialized at $\cal{X}_{\tau}$, determine the set of stable initial states\footnote{The stable initial states are the initial states from which the trajectory asymptotically converges to the origin.}, and note them as $\cal{X}_{\text{stable}} \subset \cal{X}_\tau$. The estimation of the RoA $R_i$ is given by largest sublevel set of $V_{\theta_3}$ contained in $\cal{X}$ where \eqref{eq:condition_to_fit} is satisfied, \ie,  $R_i = \cal{V}(c_i)$, where $c_i = \max_{c>0} c$ subject to
\begin{equation}
    \hat{\dot{V}}_{\theta_3}(x, u_{\theta_4}(x)) \leq -\kappa \norm{x}^2 \text{ for all } x \in \cal{X}_{\text{stable}} \intersect \cal{V}(c)
\end{equation}
 and $\cal{V}(c) \subset \cal{X}$. To enlarge the estimated RoA, we identify an exploration region by defining a level multiplier $\eta_i>1$ to include more states in the training set. The training set at iteration $i+1$ is $\cal{S}_{i+1} = \cal{V}(\eta_i c_i)$. The details are provided in Algorithm \ref{algo:max_roa}.

\subsection{Verification Condition}
The Lyapunov candidate $V_{\theta_3}$ minimizing \eqref{eq:training_objective} is called \textit{Lyapunov-like}. Define the following first-order logic formula
\begin{align} \label{eq:logic_condition}
    \Phi_\zeta (x) &= \left(\norm{x} \geq \zeta , V_{\theta_3}(x) \leq c \right) \nonumber \\
    &\wedge \left( \hat{\dot{V}}_{\theta_3}(x, u_{\theta_4}(x)) + \kappa \norm{x}^2 \geq 0 \right)
\end{align}
to check the violation of the Lyapunov condition \eqref{eq:condition_to_fit} where $\zeta >0$ is a small constant that rules out a small region around the origin to avoid numerical instabilities, and $c$ is found from Algorithm \ref{algo:max_roa}. We use dReal \cite{gao2012delta}, an SMT solver for nonlinear constraints, to solve \eqref{eq:logic_condition}. dReal runs a delta-complete algorithm whose numerical error bound is specified by the user. If dReal cannot find any solution satisfying \eqref{eq:logic_condition}, then condition \eqref{eq:condition_to_fit} is certified on $R = \cal{V}(c)$. In this case, $V_{\theta_3}$ is a CLF on $R$ for the nominal system \eqref{eq:nominal_sys} under controller $u_{\theta_4}$ by Definition~\ref{def:clf_issclf} and is an ISS-CLF for \eqref{eq:decomp_sys} by Property~\ref{thm:equiv}. Further, $R$, as a sublevel set of $V_{\theta_3}$, is forward invariant by Property~\ref{thm:forward_invariant}, and system \eqref{eq:decomp_sys} is ISS with $u_{\theta_4}$ by Property~\ref{thm:iss}. However, for computational tractability (specifically, due to the limit on the number of parameters in the dReal SMT solver that we use), we can only run the SMT solver on a simplified setting, an example for which is provided in Section \ref{sec:inv_pend}. 

Note that even if condition \eqref{eq:condition_to_fit} is violated on $R$, it is still possible for states to remain bounded under $u_{\theta_4}$. Let $\delta' = \max_{x \in R} \left[\hat{\dot{V}}_{\theta_3}(x, u_{\theta_4}(x)) + \kappa \norm{x}^2 \right]$ denote upper bound of the violation on $R$.  With $d$ defined in \eqref{eq:decomp_sys} and $\delta = \grad V_{\theta_3}(x)^\top d$, we have $\dot{V}_{\theta_3}(x, u_{\theta_4}(x)) \leq -\kappa \norm{x}^2 + \delta + \delta'$ on $R$. Let $\alpha_p$ and $\alpha_q$ be class $\cal{K}_\infty$ functions such that $\alpha_p + \alpha_q = \kappa \norm{x}^2$. If $\norm{x} \geq \alpha_q^{-1}\left(\norm{\delta + \delta'} \right)$ for all $x \in \partial R$, we have $\dot{V}_{\theta_3}(x, u_{\theta_4}(x)) \leq -\alpha_p(\norm{x})$ for all $x \in \partial R$. Let $y(x) = V_{\theta_3}(x, u_{\theta_4}(x))$ and $c$ be the corresponding level value of $R$. Then, by Nagumo's Theorem, $y(x(t)) \in [0,c]$ for $t \geq t_0$ if $y(x(t_0))\in [0,c]$. Hence, states will stay bounded.

\section{Experiments}\label{sec:experiments}
\begin{figure}[!htbp]
    \centerline{\includegraphics[width = 0.7\linewidth,clip=true,trim=0in 0.05in 0in 0.2in]{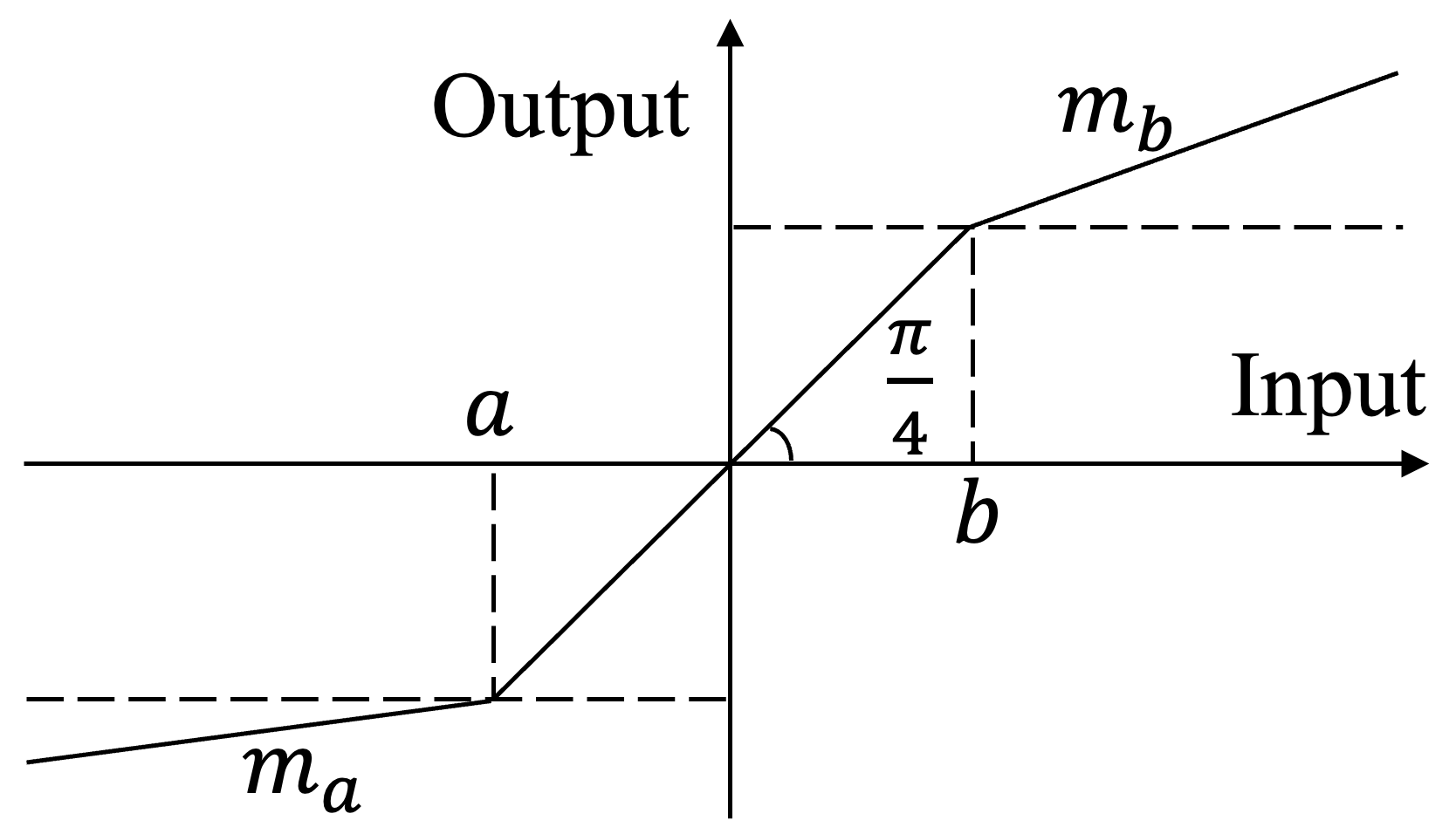}}
    \caption{Loose saturation function.}
    \label{fig:loose_saturation}
\end{figure}

In this section, we study the efficacy of our proposed learning framework. The Lyapunov candidate $V_{\theta_3}$ is randomly initialized and then pretrained to 
\begin{equation}
    V_0(x) = 0.1 x^\top x
\end{equation}
on $\cal{X}$. The controller $u_{\theta_4}$ has two parts:
\begin{equation}
    u_{\theta_4}(x) = \textrm{LS}_{a,b,m_a, m_b}( u_0(x) + \psi(x))
    \label{eq:u_theta_4}
\end{equation}
where $u_0$ is an initial locally stabilizing controller (fixed during training), $\psi$ is the trainable part, and $\textrm{LS}_{a,b,m_a, m_b}$ is the loose saturation filter (see Fig. \ref{fig:loose_saturation}). The thresholds $a$ and $b$ are fixed, but the slopes $m_a$ and $m_b$ (initialized to zero) are trainable and included in the controller parameters $\theta_4$. The motivation for using a loose saturation filter is to more clearly demonstrate that our algorithm is indeed learning to enlarge the RoA, as for some systems, a locally stable controller could automatically be a globally stable controller in the absence of such a filter. Also, a similar setup is used in reference \cite{mehrjou2021neural}, and we use the loose saturation filter in our work for a fair comparison. In \eqref{eq:u_theta_4}, $u_0$ is the linear controller given by the Linear–Quadratic Regulator (LQR) solution using the nominal dynamics (linearized at $x = 0$). Furthermore, it is to be noted that our algorithm can still work without any initial stabilizing controller. In this case, we need to ensure that the training set always has a sufficient amount of data during training, and it could take a longer time to stabilize the learning in the beginning.  

We test our approach on three examples: an inverted pendulum, a third-order strict-feedback system, and a cart-pole system. The hyperparameters and structures of neural networks are listed in Tables \ref{tab:hyperparam} and \ref{tab:nn_structure}. The experiment results are reported in Table \ref{tab:results} along with the baseline given by the estimated RoA by the LQR solution. The LQR Lyapunov function $V_{\textrm{LQR}} = x^\top P x$ is obtained using the nominal dynamics before training. Then, the estimation RoA given by $V_{\textrm{LQR}}$ is $\cal{V}(c')$, where $c' = \max_{c>0} c$ subject to $\hat{\dot{V}}_{\textrm{LQR}}(x, u_{\theta_4}(x)) \leq -\kappa \norm{x}^2$ for all $x \in \cal{X}_{\text{stable}} \intersect \cal{V}(c)$ and $\cal{V}(c) \subset \cal{X}$. Our method enlarges the true RoA by 740\%, 219\%, and 173\% on the three examples and increases the estimated RoA by at least 200\% compared with the baseline. 
For examples 1 to 3, the average training times for each iteration (as in Algorithm \ref{algo:max_roa}) are 147, 90, and 109~\si{s} respectively on an i7-5930K CPU. The sample sizes are 10000, 15625, and 10000, respectively. 

\begin{table}[!htb]
\caption{Hyperparameters used for training (a dash means that the level multiplier $\eta_i$ is kept fixed during training). For all examples, $\gamma=10^{-6}$, $\kappa=0.1$, $\epsilon=0.01$.}
\centering
\begin{threeparttable}
\begin{tabular*}{\linewidth}{c @{\extracolsep{\fill}} ccc}
\toprule\midrule
& \thead{Inverted \\ Pendulum} & \thead{Strict \\Feedback Form} & \thead{Cart-pole}\\ \midrule
$\lambda_{\text{RoA}}$  & 1000 & 500 & 500\\
$\lambda_{\text{Lip}}$  & 0.1 & 0.01 & 0.01\\
$\eta_0$  & 5 & 2 & 9\\
$k_\eta$  & 15 & - & -\\
$a$ & -2 & -1 & -5\\
$b$ & 2 & 1 & 5\\
\bottomrule\addlinespace[1ex]
\end{tabular*}
\end{threeparttable}
\label{tab:hyperparam}
\end{table}

\begin{table}[!htb]
\caption{Network structures and activation functions (id stands for the identity mapping).}
\centering
\begin{threeparttable}
\begin{tabular}{c c c c}
\toprule\midrule
& \thead{Inverted \\ Pendulum} & \thead{Strict \\Feedback Form} & \thead{Cart-pole}\\ \midrule
$f_{\theta_1}$ & \makecell{{[16,16,16,1]} \\ {[tanh,tanh,tanh,id ]}} & \makecell{{[16,16,16,3]} \\ {[tanh,tanh,tanh,id]}} & \makecell{{[16,16,16,2]} \\ {[tanh,tanh,tanh,id]}}  \\
$g_{\theta_2}$ & scalar & scalar & \makecell{{[16,16,16,2]} \\ {[tanh,tanh,tanh,id]}}  \\
\makecell{$\phi$\\ (in $V_{\theta_3}$)} & \makecell{{[64,64,64]} \\ {[tanh,tanh,tanh]}} & \makecell{{[64,64,64]} \\ {[tanh,tanh,tanh]}} & \makecell{{[64,64,64]} \\ {[tanh,tanh,tanh]}} \\
\makecell{$\psi$\\ (in $u_{\theta_4}$)} & \makecell{{[16,16,16,1]} \\ {[tanh,tanh,tanh,id]}} & \makecell{{[16,16,16,1]} \\ {[tanh,tanh,tanh,id]}} & \makecell{{[16,16,16,1]} \\ {[tanh,tanh,tanh,id]}} \\
\bottomrule\addlinespace[1ex]
\end{tabular}
\end{threeparttable}
\label{tab:nn_structure}
\end{table}

\begin{table}[!htb]
\caption{Percentages of RoA, forward invariant RoA, and estimated RoAs over the state space.}
\centering
\begin{threeparttable}
\begin{tabular}{c c c c}
\toprule\midrule
& \thead{Inverted \\ Pendulum} & \thead{Strict \\Feedback Form} & \thead{Cart-pole}\\ \midrule
\makecell{True RoA\tnote{*} \\ (before/after training)} & 11.9/100 & 31.1/99.0 & 27.8/76.0\\
\makecell{Forward invariant RoA\tnote{**} \\ (before/after training)}  & 11.9/81.7 & 30.6/70.8 & 16.7/19.2\\
Estimated RoA (ours)  & 37.4 & 16.3 & 1.2\\
Estimated RoA (LQR)  & 9.6 & 0.8 & 0.4\\
\bottomrule\addlinespace[1ex]
\end{tabular}
\begin{tablenotes}\footnotesize
\item[*] The percentages presented in Table III are approximate estimations based on a mesh. For example, the percentage of the true RoA is defined as the ratio of the number of stable mesh points over the total number of mesh points.
\item[**] Forward invariant RoA refers to the set of initial states from which the trajectory never leaves $\cal{X}$ before converging to the origin. 
\end{tablenotes}
\end{threeparttable}
\label{tab:results}
\end{table}

\subsection{Stationary Inverted Pendulum}\label{sec:inv_pend}
The stationary inverted pendulum is a second-order nonlinear system with dynamics: 
\begin{equation*}
    ml^2 \ddot{\theta} - mgl \sin \theta = u
\end{equation*}
where $\theta = 0$ is the upward position of the pendulum, and $\theta$ is positive in the counter-clockwise direction. The states are $(\theta, \omega)$ with $\omega = \dot{\theta}$ and the state space is
\begin{equation*}
    \cal{X} = [-\pi, \pi] \times [-\pi, \pi].
\end{equation*}
The true parameters of the system are 
$m = 1$ \si{kg} and $l=0.5$ \si{m}, while the nominal parameters are $m' = 0.8$ \si{kg} and $l'=0.4$ \si{m}. Thus, our initial knowledge of the system is not accurate. When determining the output dimensions of $f_{\theta_1}$ and $g_{\theta_2}$, we have taken some known relationships into account, \eg, $\dot{\theta} = \omega$ for the inverted pendulum. Therefore, for this example, the output dimensions of $f_{\theta_1}$ and $g_{\theta_2}$ should both be one. The same reasoning has been applied to the other example.

\begin{figure}[!htb]
    \centering
    \begin{subfigure}[b]{0.49\linewidth}
        \centering
        \includegraphics[width=\textwidth,clip=true,trim=0in 0.05in 0in 0.1in]{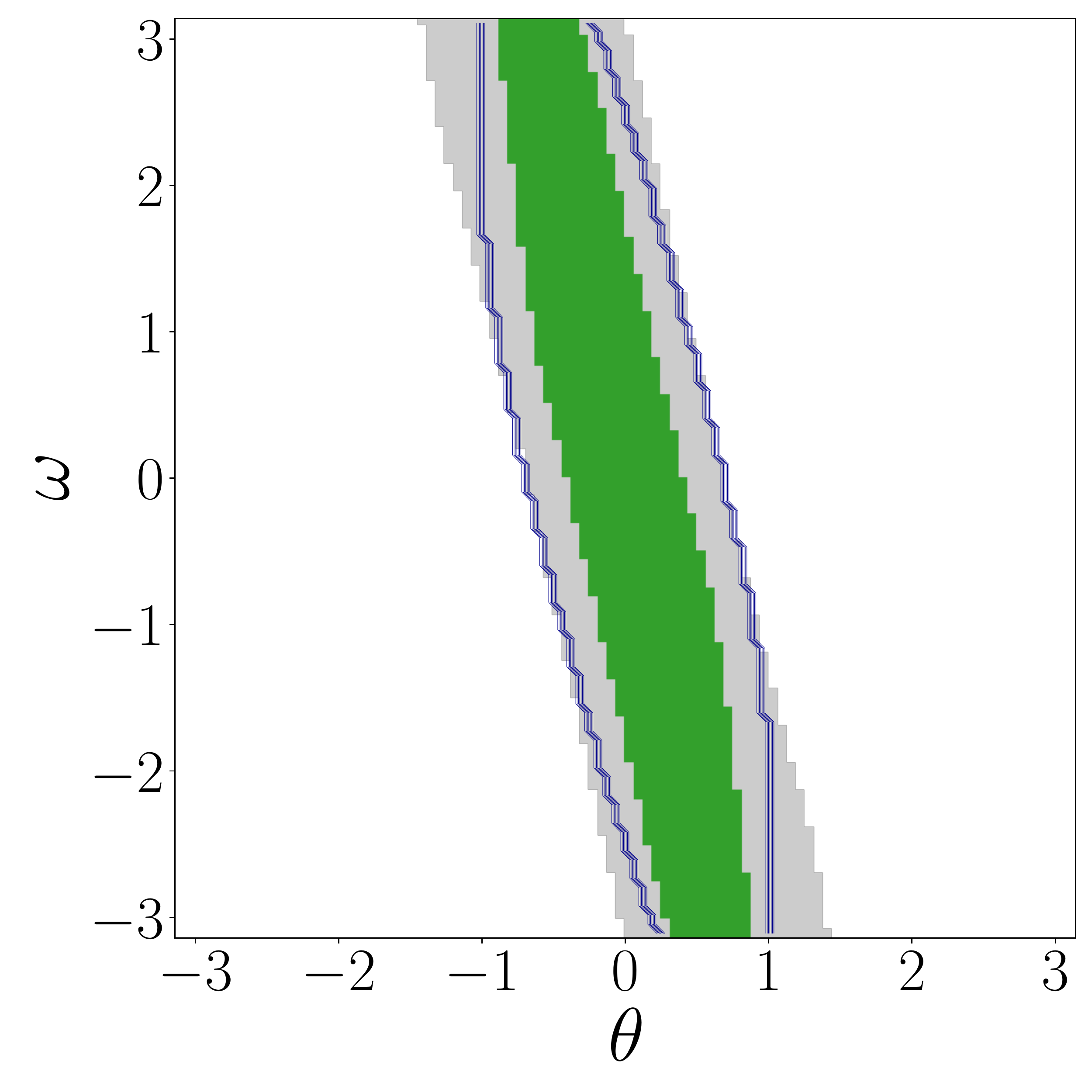}
    \end{subfigure}
    \begin{subfigure}[b]{0.49\linewidth}
        \centering
        \includegraphics[width=\textwidth,clip=true,trim=0in 0.05in 0in 0.1in]{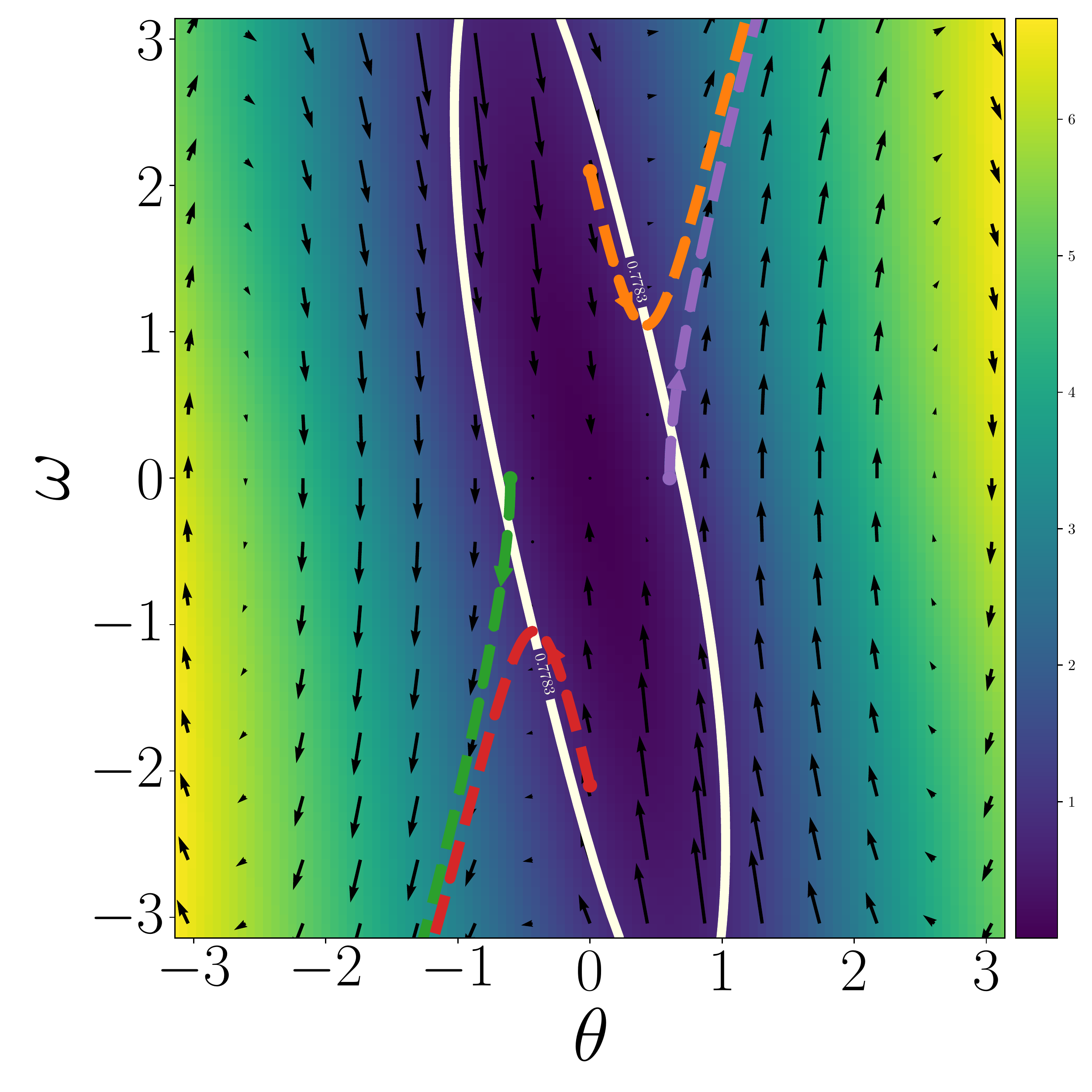}
    \end{subfigure}
    \caption{Estimated RoA by NLR \cite{mehrjou2021neural} without accounting for the dynamic uncertainty during training. Left: true RoA (green), false RoA based on the nominal model (gray), and estimated RoA (blue contour). Right: phase plot and four divergent trajectories starting from within the estimated RoA.}
    \label{fig:nlr}
\end{figure}

\begin{figure}[!htb]
    \centering
    \begin{subfigure}[b]{0.49\linewidth}
        \centering
        \includegraphics[width=\textwidth,clip=true,trim=0in 0.05in 0in 0.1in]{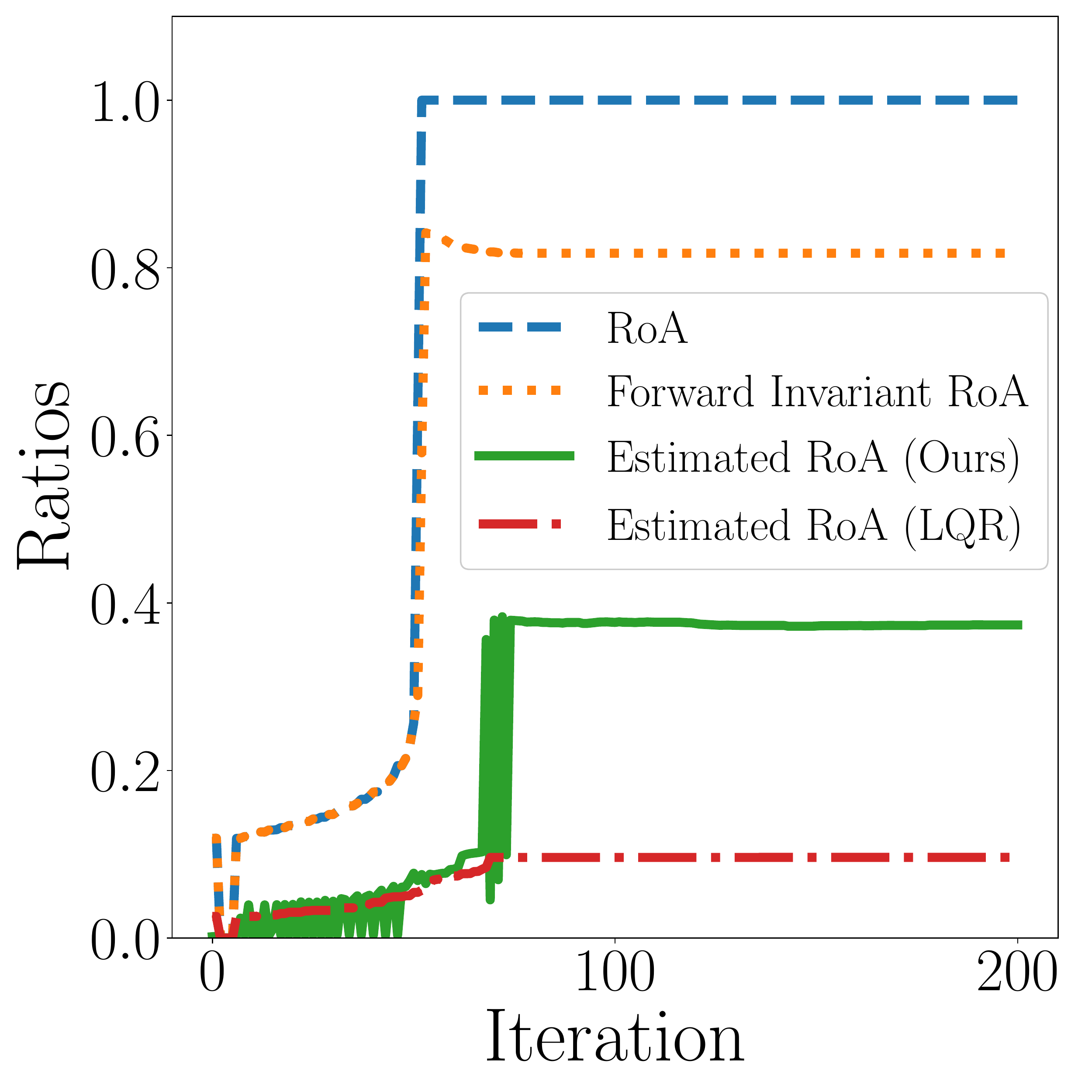}
    \end{subfigure}
    \begin{subfigure}[b]{0.49\linewidth}
        \centering
        \includegraphics[width=\textwidth,clip=true,trim=0in 0.05in 0in 0.1in]{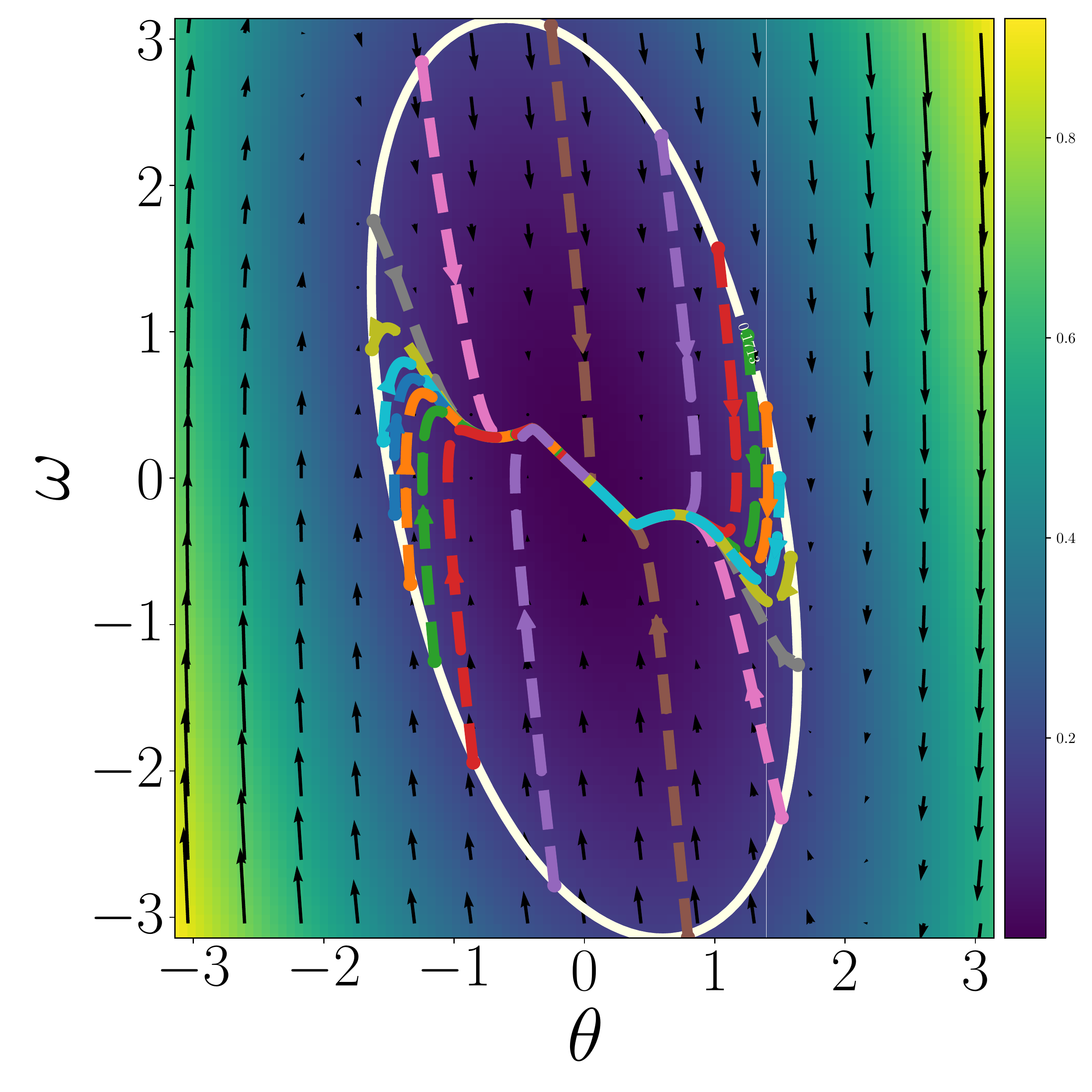}
    \end{subfigure}
    \caption{Left: true RoA and estimated RoA ratios. Right: phase plot and 20 randomly sampled trajectories starting from the boundary of the estimated RoA.}
    \label{fig:eg1_results}
\end{figure}

\begin{figure}[!htb]
    \centering
    \begin{subfigure}[b]{0.49\linewidth}
        \centering
        \includegraphics[width=\textwidth,clip=true,trim=0in 0.0in 0in 0.0in]{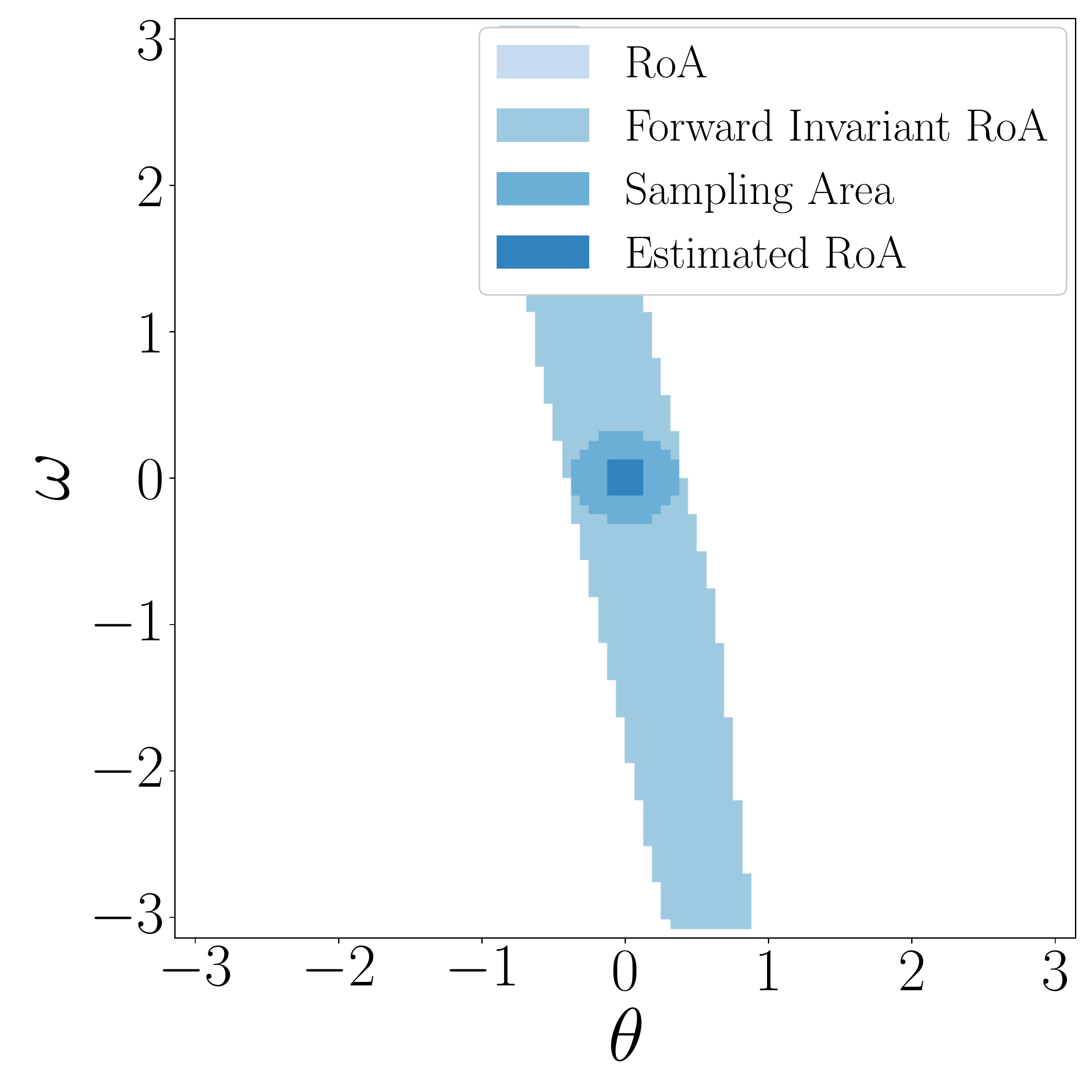}
    \end{subfigure}
    \begin{subfigure}[b]{0.49\linewidth}
        \centering
        \includegraphics[width=\textwidth,clip=true,trim=0in 0.0in 0in 0.0in]{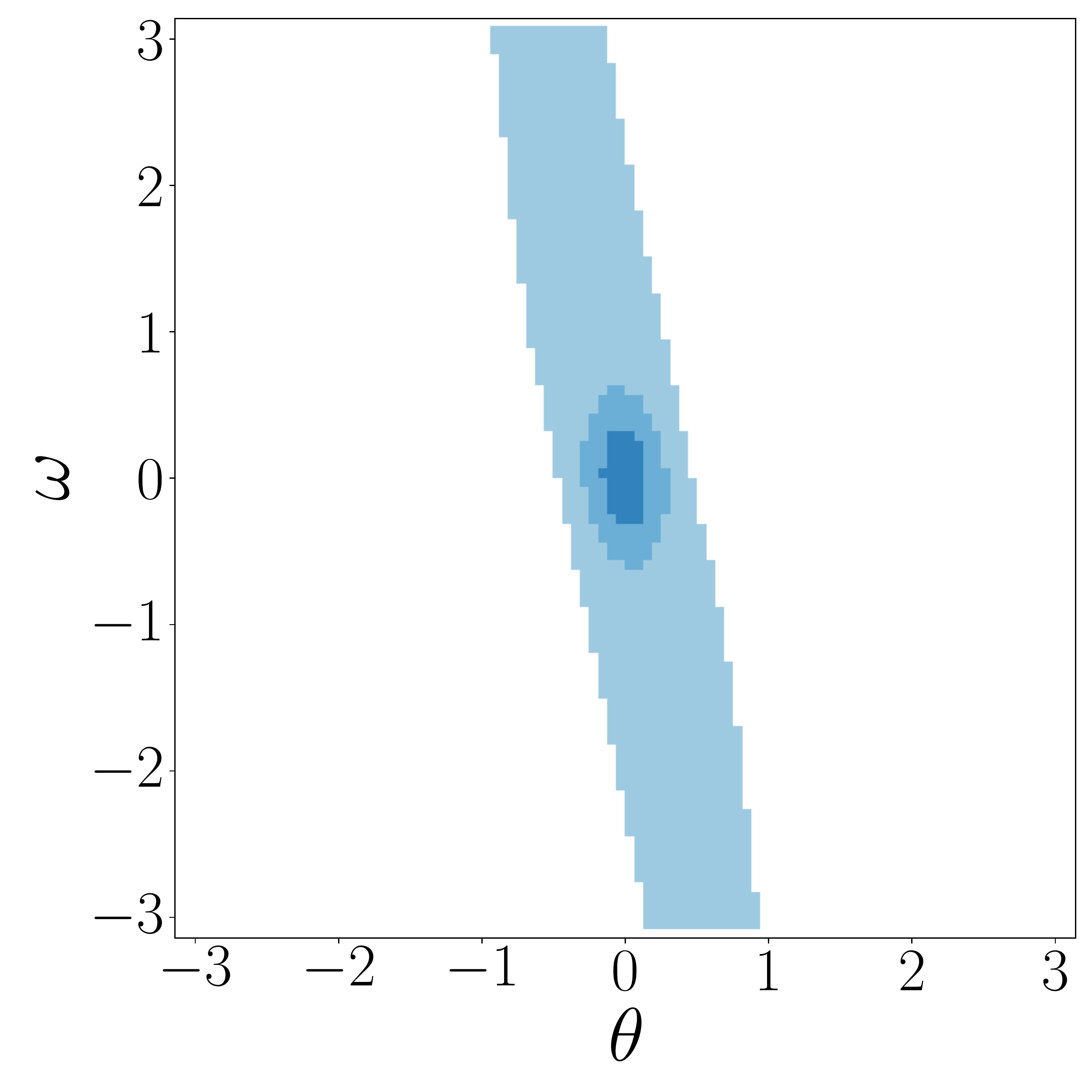}
    \end{subfigure}
    
    \begin{subfigure}[b]{0.49\linewidth}
        \centering
        \includegraphics[width=\textwidth,clip=true,trim=0in 0.0in 0in 0.0in]{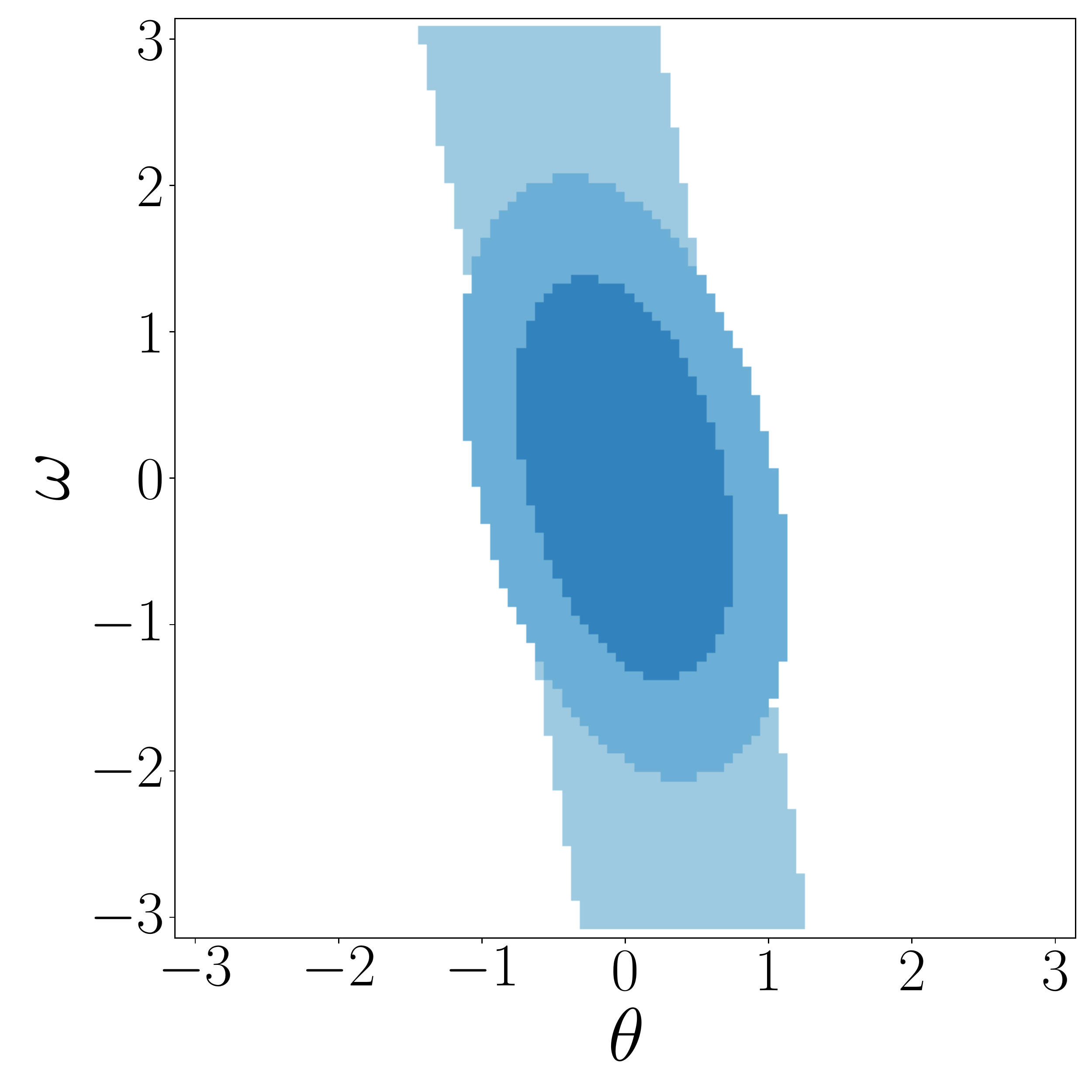}
    \end{subfigure}
    \begin{subfigure}[b]{0.49\linewidth}
        \centering
        \includegraphics[width=\textwidth,clip=true,trim=0in 0.0in 0in 0.0in]{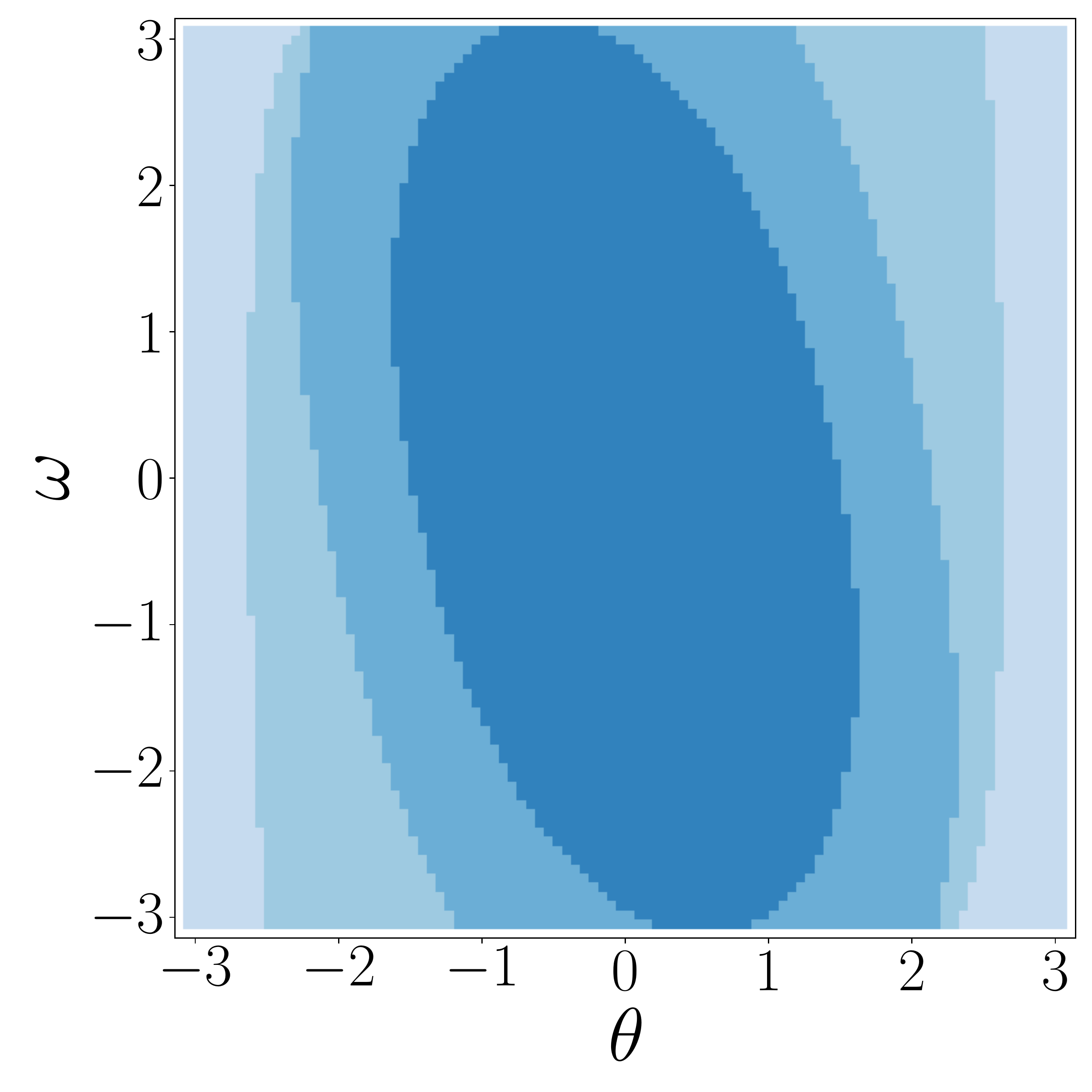}
    \end{subfigure}
    
    \begin{subfigure}[b]{0.49\linewidth}
        \centering
        \includegraphics[width=\textwidth,clip=true,trim=0in 0.0in 0in 0.0in]{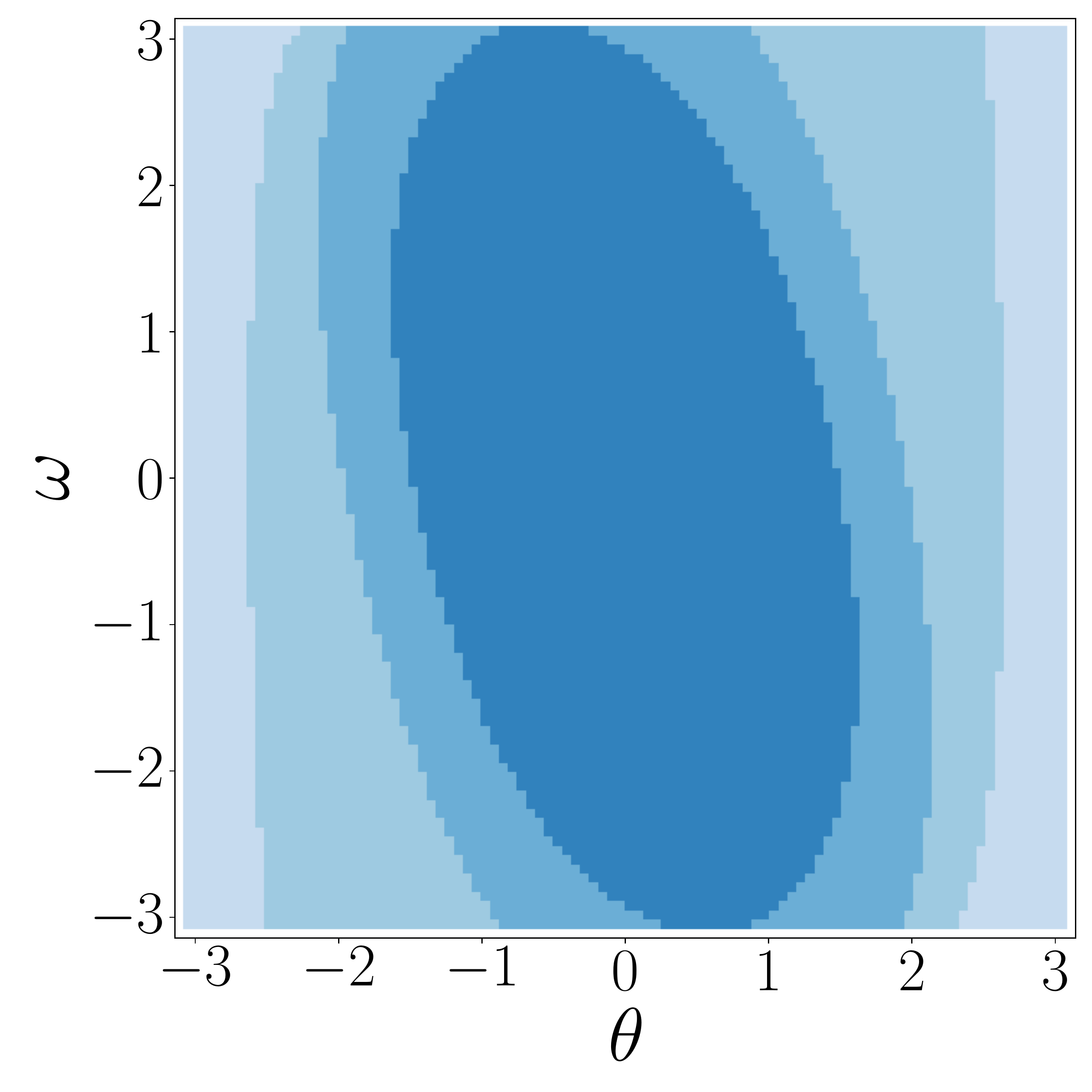}
    \end{subfigure}
    \begin{subfigure}[b]{0.49\linewidth}
        \centering
        \includegraphics[width=\textwidth,clip=true,trim=0in 0.0in 0in 0.0in]{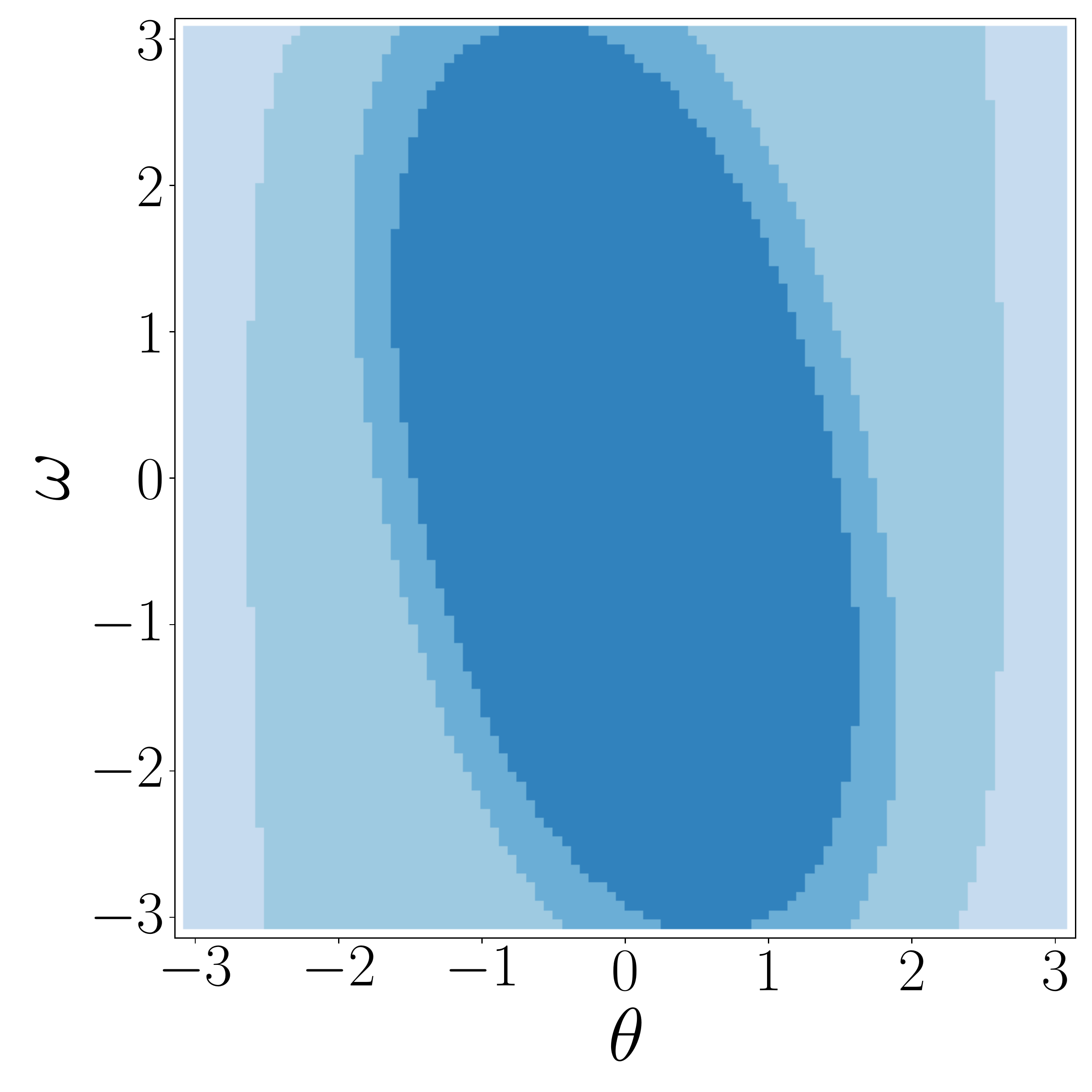}
    \end{subfigure}
    \caption{Evolution of the estimated RoA, sampling area $\cal{S}_i$, forward invariant RoA, and true RoA (see legend in the first figure). The number of iterations is 0, 25, 50, 75, 100, and 200 from top left to bottom right. 
    }
    \label{fig:eg1_roas}
\end{figure}

Figure \ref{fig:nlr} demonstrates an overestimation of the true RoA generated by the offline method called Neural Lyapunov Redesign (NLR, \cite{mehrjou2021neural}), which does not consider uncertainties in the system dynamics. Figure \ref{fig:eg1_results} shows the percentages of RoA, forward invariant RoA, and estimated RoA over the entire state space. The enlargement of the true RoA and the estimated RoA based on our method are reflected in Fig. \ref{fig:eg1_roas}.

Next, we consider using an SMT solver to verify the CLF. Due to the SMT solver's limit on the number of parameters, we consider a simpler setting for \eqref{eq:logic_condition}. We reduce $f_{\theta_1}$, $\phi$, and $\psi$ to neural networks of three, two, and three layers, respectively. The nonlinear constraint \eqref{eq:logic_condition} is solved with $\zeta = 0.3$ and a precision of $10^{-3}$, and no counter-example is found. This means that \eqref{eq:condition_to_fit} is verified on the estimated RoA, which is around $34.6\%$.

\subsection{A Third-Order Strict Feedback Form} \label{sec:backstepping}
\begin{figure}[!htb]
    \centering
    \begin{subfigure}[b]{0.49\linewidth}
        \centering
        \includegraphics[width=\textwidth,clip=true,trim=0in 0.05in 0in 0.1in]{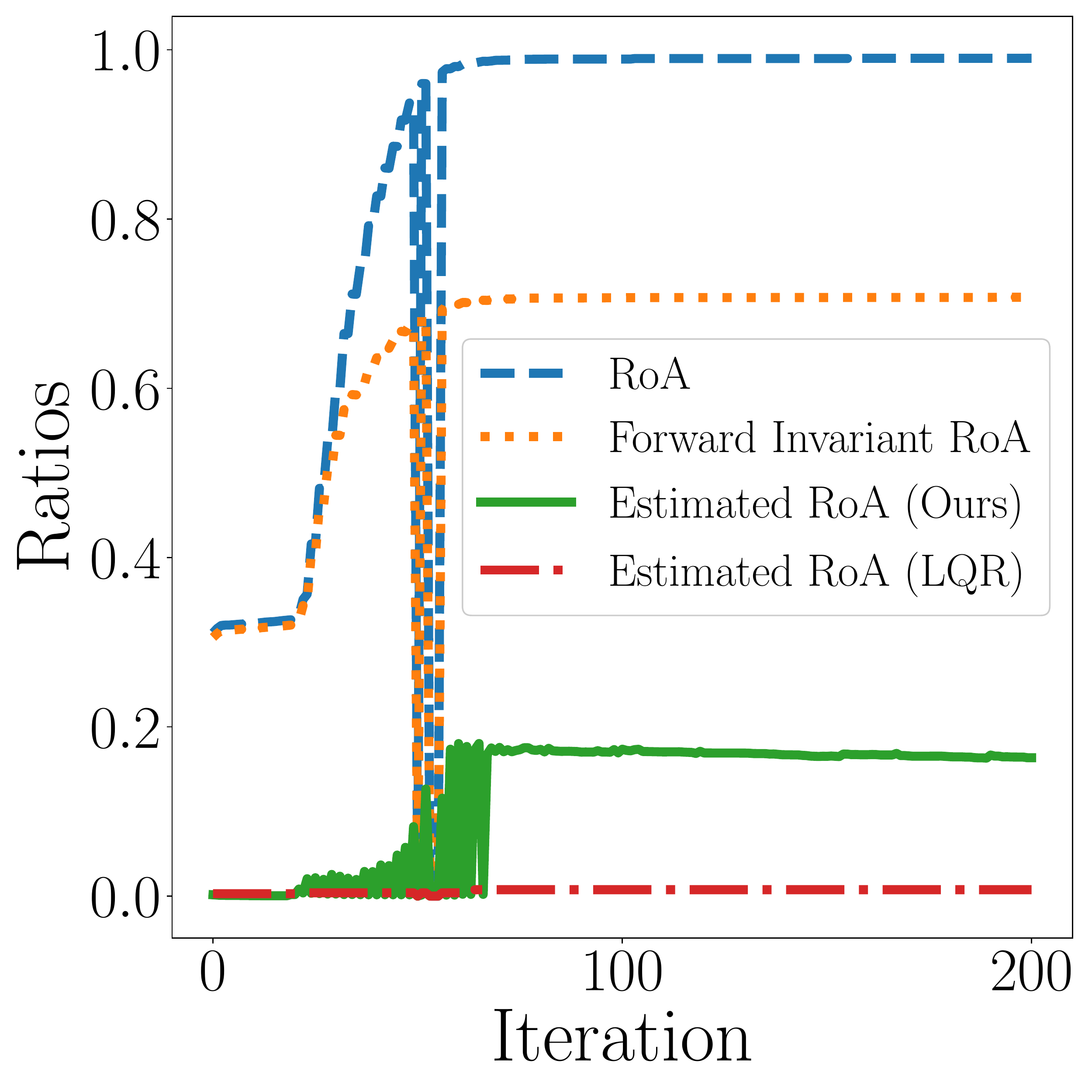}
    \end{subfigure}
    \begin{subfigure}[b]{0.49\linewidth}
        \centering
        \includegraphics[width=\textwidth,clip=true,trim=0in 0.05in 0in 0.1in]{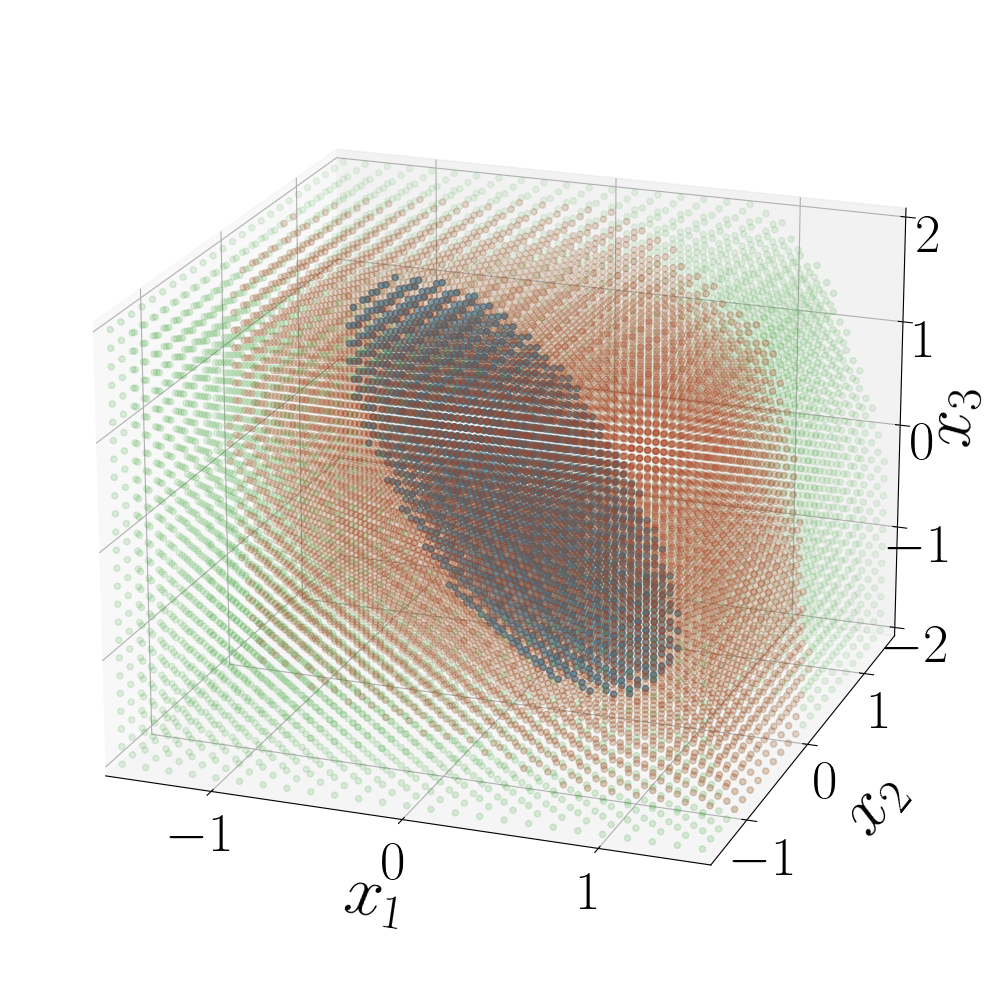}
    \end{subfigure}
    \caption{Left: true RoA and estimated RoA ratios. Right: true RoA (green), forward invariant RoA (red), and estimated RoA (blue).}
    \label{fig:eg2_results}
\end{figure}

\begin{figure}[!htb]
    \centering
    \begin{subfigure}[b]{0.49\linewidth}
        \centering
        \includegraphics[width=\textwidth,clip=true,trim=0in 0.05in 0in 0.1in]{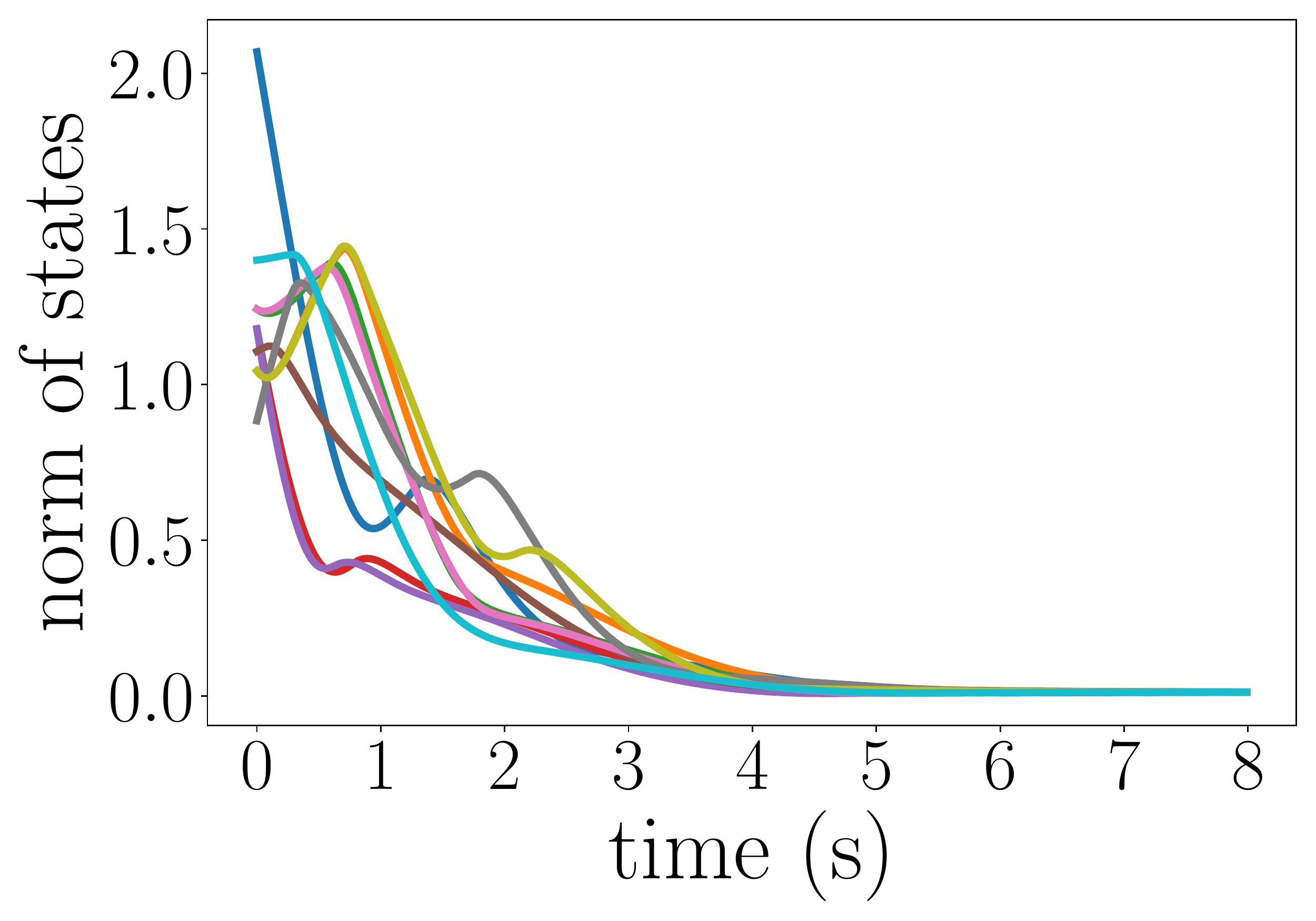}
    \end{subfigure}
    \begin{subfigure}[b]{0.49\linewidth}
        \centering
        \includegraphics[width=\textwidth,clip=true,trim=0in 0.05in 0in 0.1in]{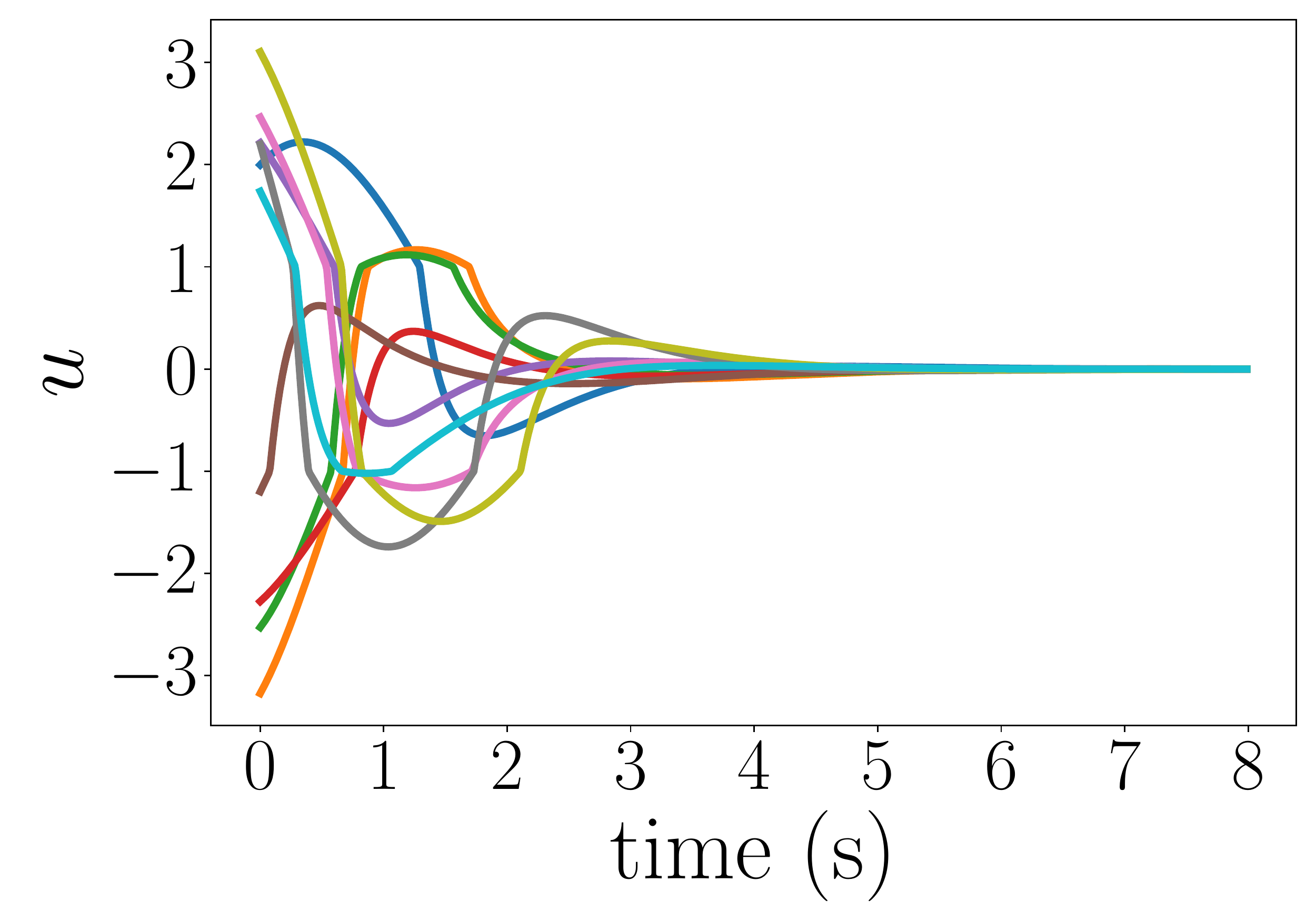}
    \end{subfigure}
    \caption{Ten randomly sampled trajectories starting from the boundary of the estimated RoA. }
    \label{fig:eg2_trajs}
\end{figure}

Consider a third-order system of the strict feedback form:
\begin{equation*}
    \dot{x}_1= e_1 x_2,\quad \dot{x}_2 = e_2 x_3,\quad \dot{x}_3 = e_3 x_1^2 + e_4 u.
\end{equation*}
The states are $(x_1, x_2, x_3)$ and the state space is 
\begin{equation*}
    \cal{X} = \Set{(x_1, x_2, x_3)}{\abs{x_1} \leq 1.5, \abs{x_2} \leq 1.5, \abs{x_3} \leq 2}.
\end{equation*}
Again, we assume that our initial knowledge of the system is not accurate: the true parameters are $e_1=1$, $e_2=1$, $e_3=1$, and $e_4=1$, while the nominal parameters are $e_1'=0.9$, $e_2'=0.8$, $e_3'=0.9$, and $e_4'=0.8$.
Figure \ref{fig:eg2_results} shows the evolution of the percentages of RoA, forward invariant RoA, and estimated RoA along with a 3D visualization. We note that as our learning framework tries to stabilize more states, there can be transition periods where the size of the RoA drops, which might result from the sensitivity of the nonlinear controller to its weights. Figure \ref{fig:eg2_trajs} demonstrates ten randomly sampled trajectories starting from the boundary of the estimated RoA.

\subsection{Cart-Pole System}\label{sec:cart-pole}
\begin{figure}[!htb]
    \centering
    \includegraphics[width=0.5\linewidth]{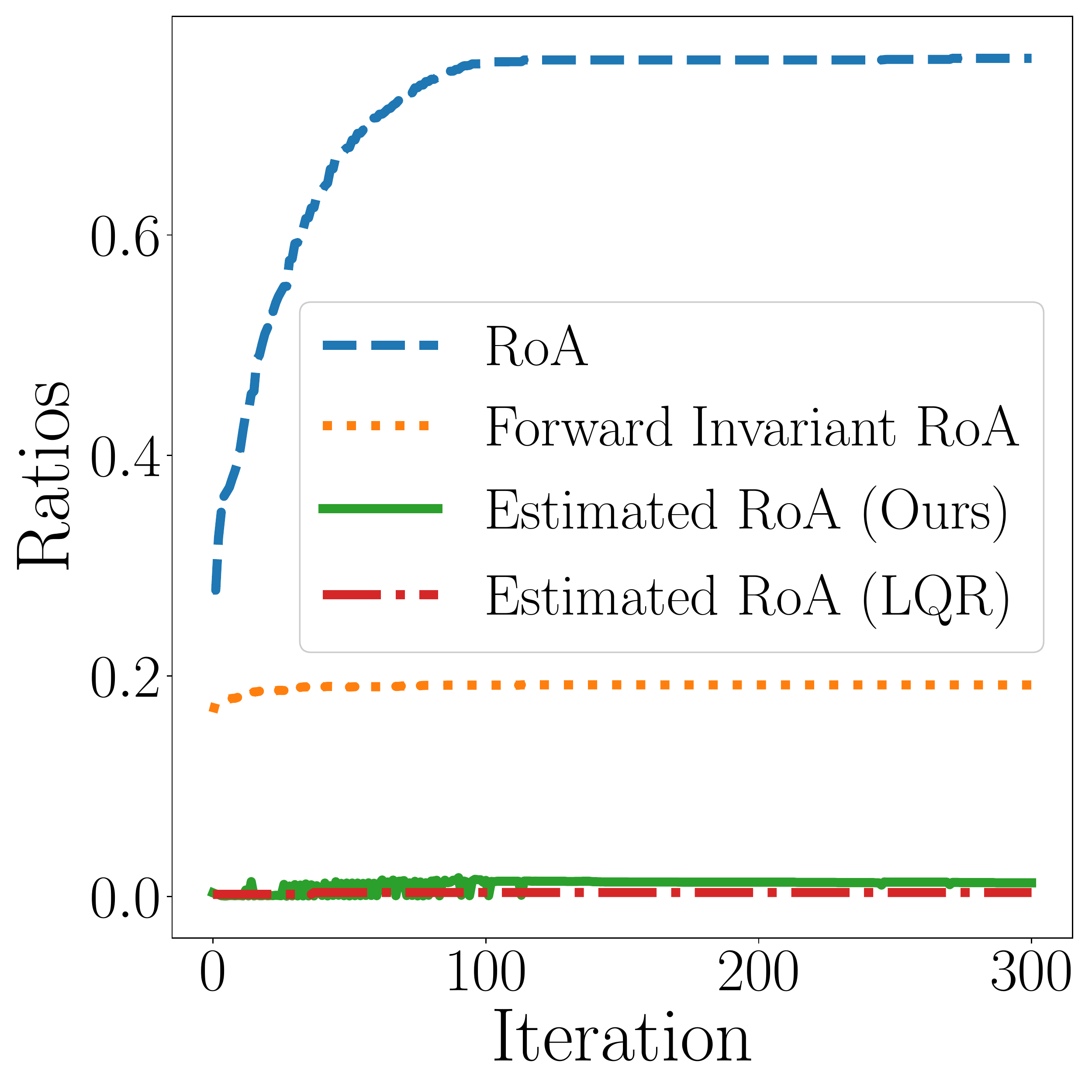}
    \caption{evolution of the percentages of RoA, forward invariant RoA, and estimated RoA. }
    \label{fig:eg3_roas}
\end{figure}

\begin{figure}[!htb]
    \centering
    \begin{subfigure}[b]{0.49\linewidth}
        \centering
        \includegraphics[width=\textwidth,clip=true,trim=0in 0.05in 0in 0.1in]{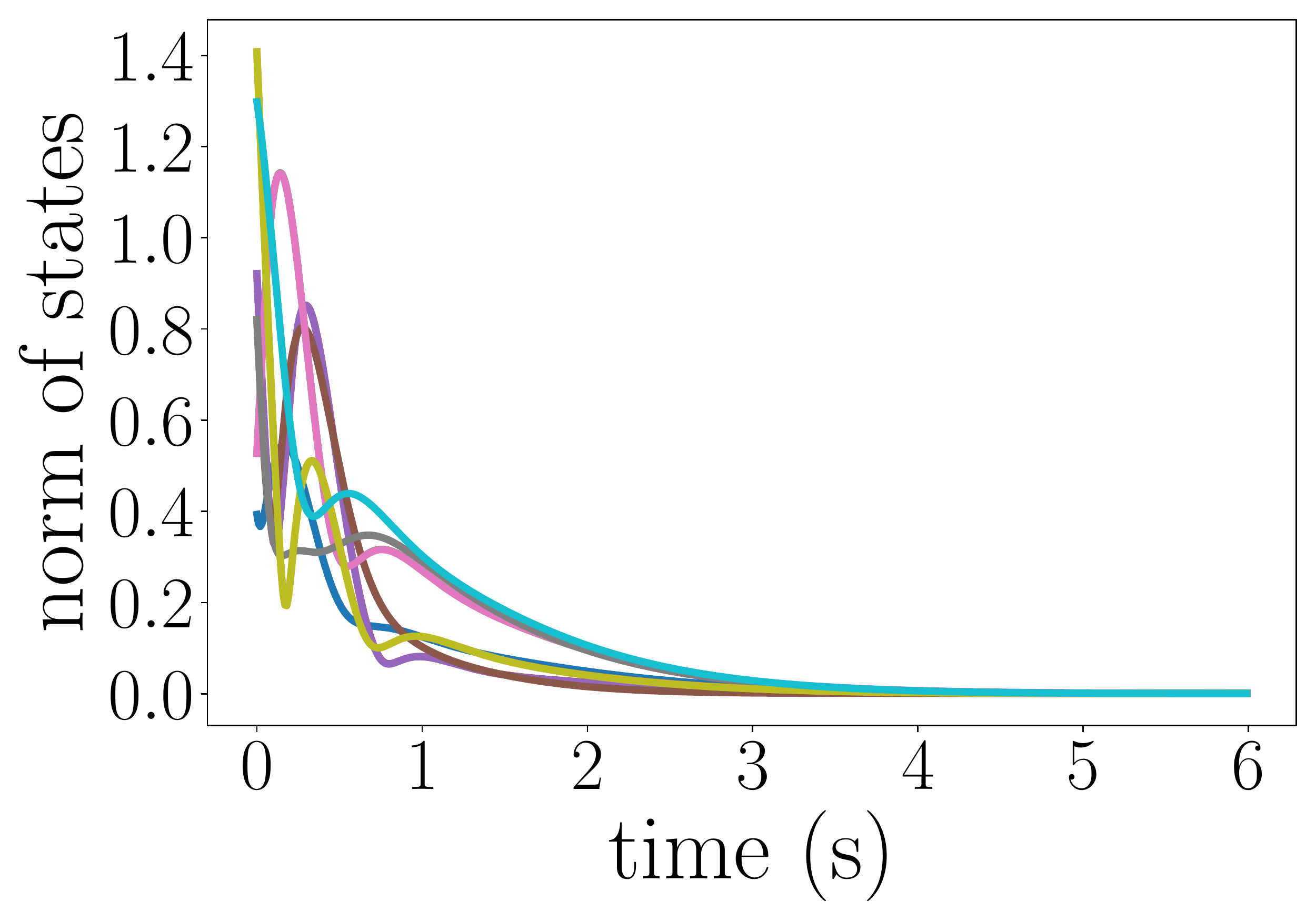}
    \end{subfigure}
    \begin{subfigure}[b]{0.49\linewidth}
        \centering
        \includegraphics[width=\textwidth,clip=true,trim=0in 0.05in 0in 0.1in]{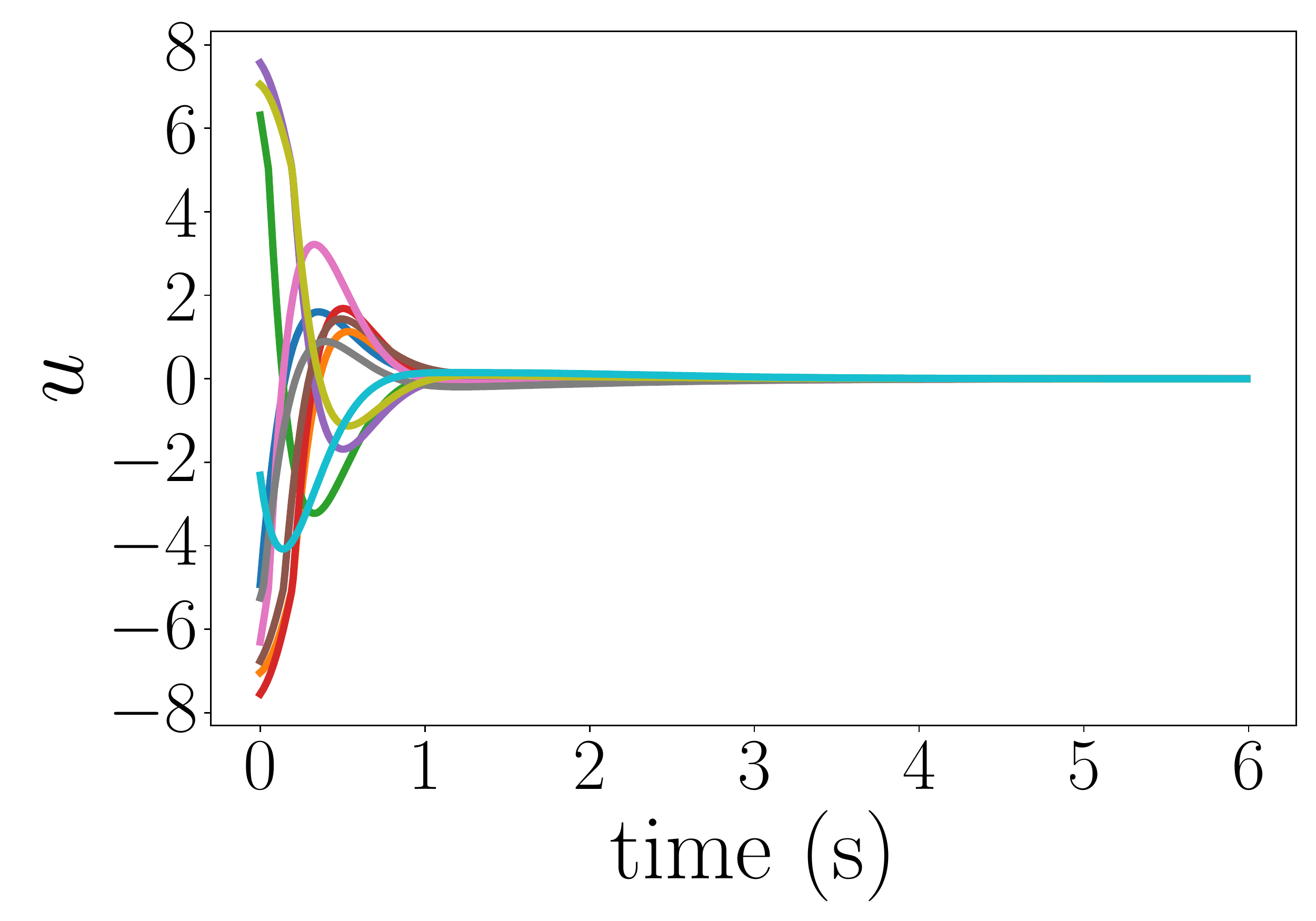}
    \end{subfigure}
    \caption{Ten randomly sampled trajectories starting from the boundary of the estimated RoA.}
    \label{fig:eg3_trajs}
\end{figure}

\begin{figure}[!htb]
    \centering
    \begin{subfigure}[b]{0.49\linewidth}
        \centering
        \includegraphics[width=\textwidth,clip=true,trim=0.0in 0.05in 0in 0.1in]{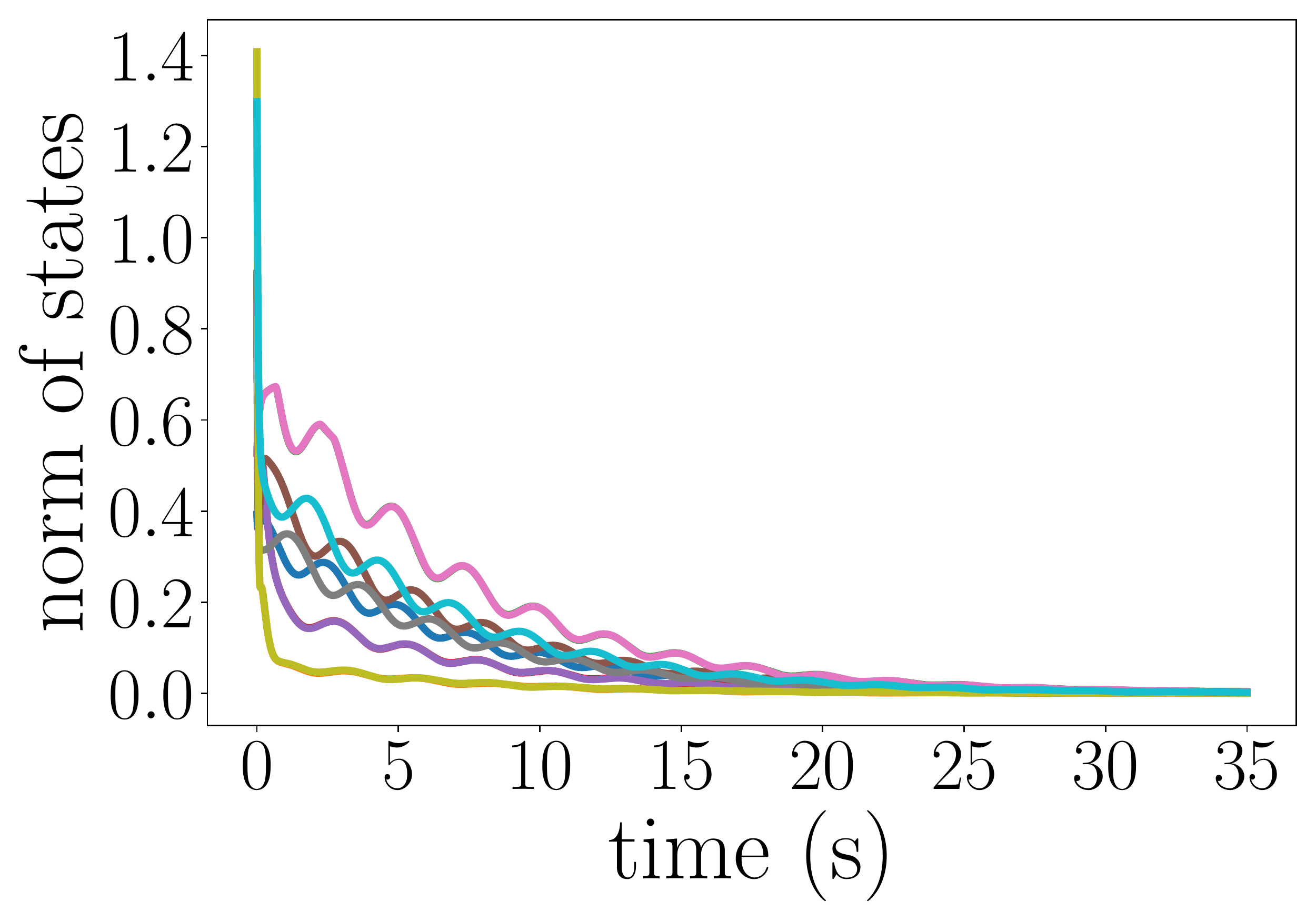}
    \end{subfigure}
    \begin{subfigure}[b]{0.49\linewidth}
        \centering
        \includegraphics[width=\textwidth,clip=true,trim=0.0in 0.05in 0in 0.1in]{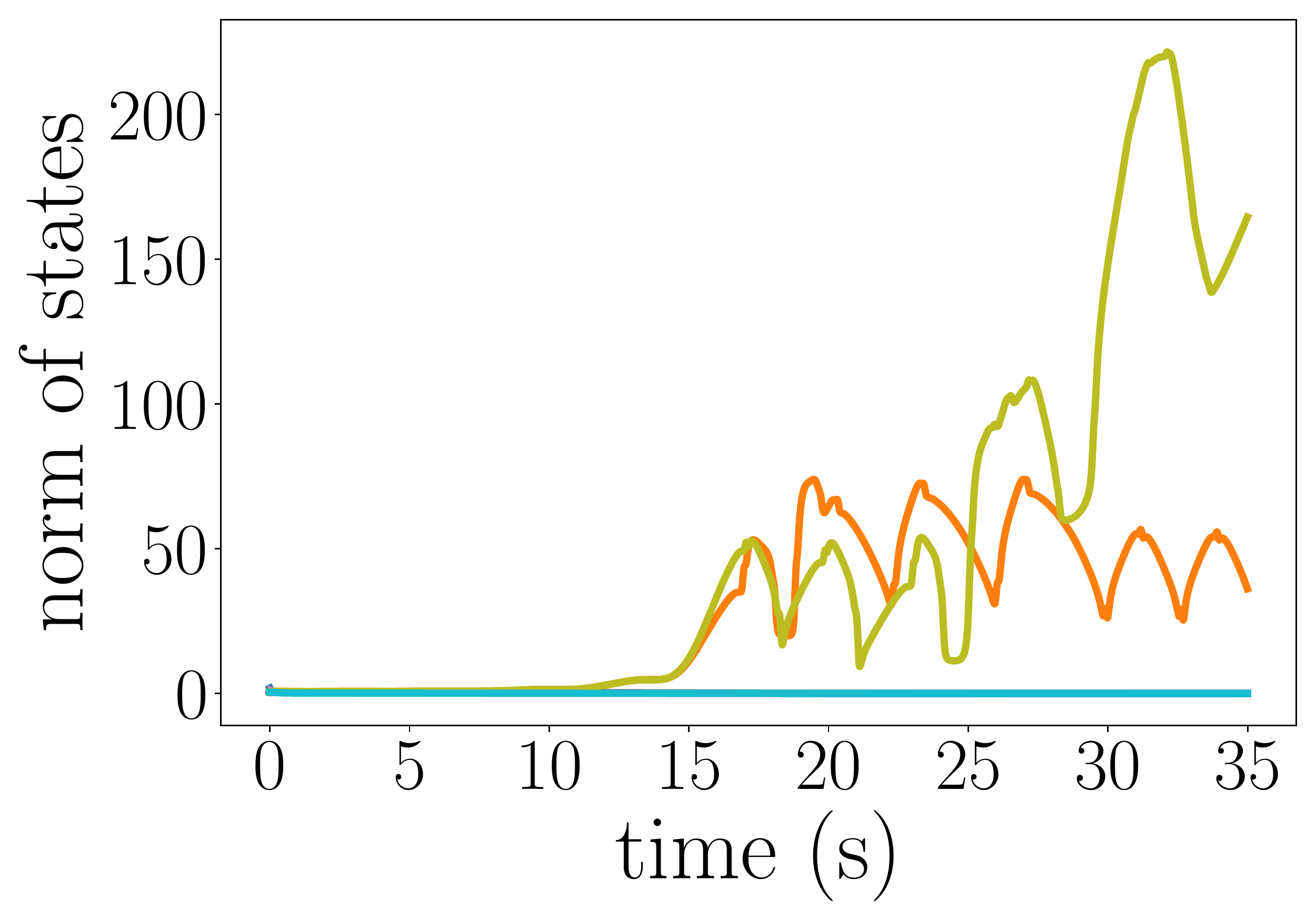}
    \end{subfigure}
    \caption{Ten randomly sampled trajectories (some visually overlapped in the plots) starting from boundaries of RoAs estimated by our method (left) and NLR (right), respectively.}
    \label{fig:eg3_perturb}
\end{figure}

Finally, our method is tested on the cart-pole system: 
\begin{align*}
    (M+m)\ddot{x} - ml \ddot{\theta} \cos{\theta} + ml \dot{\theta}^2 \sin{\theta} + b_c \dot{x} &= u,\\
    ml^2 \ddot{\theta} - mgl \sin{\theta} &= ml\ddot{x} \cos{\theta}.
\end{align*}
The states are $(\theta, \omega, x, v)$, where $\omega = \dot{\theta}$ and $v = \dot{x}$. The state space is defined as 
\begin{equation*}
    \cal{X} = \Set{(\theta, \omega, x, v)}{\abs{\theta} \leq \pi/6, \abs{\omega} \leq 1, \abs{x} \leq 1, \abs{v} \leq 1.5}.
\end{equation*}
Note that $\theta = 0$ corresponds to the upward position of the pole, and $\theta$ is positive in the counter-clockwise direction. The true parameters are $M = 1$ \si{kg}, $m = 0.3$ \si{kg}, $l = 1$ \si{m}, and $b_c = 0$ \si{kg.s^{-1}}, while the nominal parameters are $M' = 0.8$ \si{kg}, $m' = 0.27$ \si{kg}, $l' = 0.8$ \si{m}, and $b_c' = 0$ \si{kg.s^{-1}}.
Figure \ref{fig:eg3_roas} shows the evolution of the percentages of RoA, forward invariant RoA, and estimated RoA. Ten randomly sampled trajectories starting from the boundary of the estimated RoA are plotted in Fig. \ref{fig:eg3_trajs}. The offline method NLR (\cite{mehrjou2021neural}) generates an estimated forward invariant RoA ratio of 1.7\% using the true parameters, which demonstrates the difficulty of this task. Finally, we test the robustness of the RoA estimation to sudden perturbations, where $b_c$ is increased to $9.1$ \si{kg.s^{-1}}. As seen in Fig. \ref{fig:eg3_perturb}, our estimated RoA remains valid while the NLR estimated RoA no longer holds.

\vspace*{-0.3cm}

\section{Conclusion} \label{sec:conclusion}
We proposed a learning framework that can synthesize state-feedback controllers and CLF for control-affine nonlinear systems with unstructured uncertainties. Our approach initializes the system at different initial conditions and observes the system trajectories. Exact knowledge of system dynamics is not required. Based on a regularity condition, we model uncertainties as bounded and structured disturbances. Experiments show that our method can find a controller that enlarges the RoA and a CLF that estimates the RoA. In addition, our work produces RoA estimations with the uncertainties considered to avoid overestimation (Section \ref{sec:inv_pend}) and has better robustness to sudden changes in dynamics (Section \ref{sec:cart-pole}). Future work includes relaxing assumptions, generalizing our method to higher-dimension systems, and exploring further combinations of the proposed approach and neural network verifying tools (such as SMT solvers). 




\bibliographystyle{IEEEtran}
\bibliography{master}

\begin{thebibliography}{10}
\providecommand{\url}[1]{#1}
\csname url@samestyle\endcsname
\providecommand{\newblock}{\relax}
\providecommand{\bibinfo}[2]{#2}
\providecommand{\BIBentrySTDinterwordspacing}{\spaceskip=0pt\relax}
\providecommand{\BIBentryALTinterwordstretchfactor}{4}
\providecommand{\BIBentryALTinterwordspacing}{\spaceskip=\fontdimen2\font plus
\BIBentryALTinterwordstretchfactor\fontdimen3\font minus
  \fontdimen4\font\relax}
\providecommand{\BIBforeignlanguage}[2]{{%
\expandafter\ifx\csname l@#1\endcsname\relax
\typeout{** WARNING: IEEEtran.bst: No hyphenation pattern has been}%
\typeout{** loaded for the language `#1'. Using the pattern for}%
\typeout{** the default language instead.}%
\else
\language=\csname l@#1\endcsname
\fi
#2}}
\providecommand{\BIBdecl}{\relax}
\BIBdecl

\bibitem{sontag1989universal}
E.~D. Sontag, ``A `universal'construction of {Artstein's} theorem on nonlinear
  stabilization,'' \emph{Systems \& Control Lett.}, vol.~13, no.~2, pp.
  117--123, 1989.

\bibitem{richards2018lyapunov}
S.~M. Richards, F.~Berkenkamp, and A.~Krause, ``The {Lyapunov} neural network:
  adaptive stability certification for safe learning of dynamical systems,'' in
  \emph{Proc. Conf. on Robot Learning}, (Z{\"u}rich, Switzerland), Oct. 2018,
  pp. 466--476.

\bibitem{dai2021state}
B.~Dai, P.~Krishnamurthy, A.~Papanicolaou, and F.~Khorrami, ``State constrained
  stochastic optimal control using {LSTMs},'' in \emph{Proc. American Control
  Conf.}, (Virtual), May 2021, pp. 1294--1299.

\bibitem{DBLP:conf/cdc/DaiKK22}
B.~Dai, P.~Krishnamurthy, and F.~Khorrami, ``Learning a better control barrier
  function,'' in \emph{Proc. IEEE Conf. on Decision and Control}, (Cancun,
  Mexico), Dec. 2022, pp. 945--950.

\bibitem{mehrjou2021neural}
A.~Mehrjou, M.~Ghavamzadeh, and B.~Sch{\"o}lkopf, ``Neural {Lyapunov}
  redesign,'' in \emph{Proc. Annual Learning for Dynamics \& Control Conf.},
  (Virtual), June 2021, pp. 459--470.

\bibitem{wei2020towards}
S.~Wei, X.~Chen, X.~Zhang, and C.~Qi, ``Towards safe and socially compliant
  map-less navigation by leveraging prior demonstrations,'' in \emph{Proc.
  International Conf. on Intelligent Robotics and Applications}, (Virtual), Nov
  2020, pp. 133--145.

\bibitem{lyapunov1992general}
A.~M. Lyapunov, ``The general problem of the stability of motion,''
  \emph{International J. Control}, vol.~55, no.~3, pp. 531--534, 1992.

\bibitem{artstein1983stabilization}
Z.~Artstein, ``Stabilization with relaxed controls,'' \emph{Nonlinear Analysis:
  Theory Methods \& Appl.}, vol.~7, no.~11, pp. 1163--1173, 1983.

\bibitem{sontag1995characterizations}
E.~D. Sontag and Y.~Wang, ``On characterizations of input-to-state stability
  with respect to compact sets,'' in \emph{Nonlinear Control Systems Design
  1995}.\hskip 1em plus 0.5em minus 0.4em\relax Pergamon, 1995, pp. 203--208.

\bibitem{KK06}
P.~Krishnamurthy and F.~Khorrami, ``On uniform solvability of
  parameter‐dependent {Lyapunov} inequalities and applications to various
  problems,'' \emph{SIAM Journal on Control and Optimization}, vol.~45, no.~4,
  pp. 1147--1164, 2006.

\bibitem{krstic1995control}
M.~Krsti{\'c} and P.~V. Kokotovi{\'c}, ``Control {Lyapunov} functions for
  adaptive nonlinear stabilization,'' \emph{Systems \& Control Lett.}, vol.~26,
  no.~1, pp. 17--23, 1995.

\bibitem{berkenkamp2016safe}
F.~Berkenkamp, R.~Moriconi, A.~P. Schoellig, and A.~Krause, ``Safe learning of
  regions of attraction for uncertain, nonlinear systems with {Gaussian}
  processes,'' in \emph{Proc. IEEE Conf. on Decision and Control}, (Las Vegas,
  NV), Dec. 2016, pp. 4661--4666.

\bibitem{sun2021lyapunov}
R.~Sun, M.~L. Greene, D.~M. Le, Z.~I. Bell, G.~Chowdhary, and W.~E. Dixon,
  ``{Lyapunov}-based real-time and iterative adjustment of deep neural
  networks,'' \emph{IEEE Control Syst. Lett.}, vol.~6, pp. 193--198, 2021.

\bibitem{taylor2019control}
A.~J. Taylor, V.~D. Dorobantu, M.~Krishnamoorthy, H.~M. Le, Y.~Yue, and A.~D.
  Ames, ``A control {Lyapunov} perspective on episodic learning via projection
  to state stability,'' in \emph{Proc. IEEE Conf. on Decision and Control},
  (Nice, France), Dec. 2019, pp. 1448--1455.

\bibitem{tibken2000estimation}
B.~Tibken, ``Estimation of the domain of attraction for polynomial systems via
  {LMI}s,'' in \emph{Proc. IEEE Conf. on Decision and Control}, (Sydney,
  Australia), Dec. 2000, pp. 3860--3864.

\bibitem{papachristodoulou2002construction}
A.~Papachristodoulou and S.~Prajna, ``On the construction of {Lyapunov}
  functions using the sum of squares decomposition,'' in \emph{Proc. IEEE Conf.
  on Decision and Control}, (Las Vegas, Nevada), Dec. 2002, pp. 3482--3487.

\bibitem{giesl2015review}
P.~Giesl and S.~Hafstein, ``Review on computational methods for {Lyapunov}
  functions,'' \emph{Discrete and Continuous Dynamical Systems - B}, vol.~20,
  no.~8, pp. 2291--2331, 2015.

\bibitem{chang2019neural}
Y.-C. Chang, N.~Roohi, and S.~Gao, ``Neural {Lyapunov} control,'' in
  \emph{Proc. Conf. on Neural Information Processing Systems}, (Vancouver,
  Canada), Dec. 2019.

\bibitem{zhao2021neural}
T.~Zhao, J.~Wang, X.~Lu, and Y.~Du, ``Neural {Lyapunov} control for power
  system transient stability: A deep learning-based approach,'' \emph{IEEE
  Trans. on Power Systems}, vol.~37, no.~2, pp. 955--966, 2021.

\bibitem{sontag1989smooth}
E.~D. Sontag \emph{et~al.}, ``Smooth stabilization implies coprime
  factorization,'' \emph{IEEE Trans. Autom. Control}, vol.~34, no.~4, pp.
  435--443, 1989.

\bibitem{gao2012delta}
S.~Gao, J.~Avigad, and E.~M. Clarke, ``$\delta$-complete decision procedures
  for satisfiability over the reals,'' in \emph{International Joint Conf. on
  Automated Reasoning}, (Manchester, UK), June 2012, pp. 286--300.

\end{thebibliography}

\end{document}